\documentclass{aa}
\pdfoutput=1 
\usepackage{bm} 
\usepackage[varg]{txfonts}
\usepackage{graphicx}
\usepackage{chemformula}

\usepackage[markup=bfit]{changes}
\definechangesauthor[name={Bibiana Prinoth}, color=purple]{BP}
\definechangesauthor[name={Martina Baratella}, color=teal]{MB}
\definechangesauthor[name={Julia Seidel}, color=orange]{JVS}
\definechangesauthor[name={Jens},color=blue]{HJH}
\definechangesauthor[name={Notes}, color=orange]{N}

\usepackage{booktabs}

\usepackage{natbib,twoopt}
\usepackage[breaklinks]{hyperref}      
\bibpunct{(}{)}{;}{a}{}{,}             
\definecolor{cobalt}{rgb}{0.06, 0.2, 0.65}
\hypersetup{
  colorlinks,
  citecolor=cobalt,
  linkcolor=[rgb]{0.8, 0.2, 1.0},
  urlcolor=cobalt,
}
\makeatletter
  \newcommandtwoopt{\citeads}[3][][]{\href{http://adsabs.harvard.edu/abs/#3}%
    {\def\hyper@linkstart##1##2{}%
     \let\hyper@linkend\@empty\citealp[#1][#2]{#3}}}
  \newcommandtwoopt{\citepads}[3][][]{\href{http://adsabs.harvard.edu/abs/#3}%
    {\def\hyper@linkstart##1##2{}%
     \let\hyper@linkend\@empty\citep[#1][#2]{#3}}}
  \newcommandtwoopt{\citetads}[3][][]{\href{http://adsabs.harvard.edu/abs/#3}%
    {\def\hyper@linkstart##1##2{}%
     \let\hyper@linkend\@empty\citet[#1][#2]{#3}}}
  \newcommandtwoopt{\citeyearads}[3][][]%
    {\href{http://adsabs.harvard.edu/abs/#3}
    {\def\hyper@linkstart##1##2{}%
     \let\hyper@linkend\@empty\citeyear[#1][#2]{#3}}}
\makeatother

\makeatletter
\renewcommand*\aa@pageof{, page \thepage{} of \pageref*{LastPage}}

\usepackage[separate-uncertainty = true,multi-part-units=single, group-digits = integer, group-four-digits = true,per-mode=reciprocal]{siunitx}

\newcommand{\numpm}[3]{$#1^{#2}_{#3}$}
\newcommand{\target}{WASP-121\,b\xspace}
\newcommand{\shone}{\texttt{shone}\xspace}

\newcommand{\alldetectionsccf}{\ion{H}{I}, \ion{Li}{I}, \ion{Na}{I}, \ion{Mg}{I}, \ion{K}{I}, \ion{Ca}{I}, \ion{Ti}{I}, \ion{V}{I}, \ion{Cr}{I}, \ion{Mn}{I}, \ion{Fe}{I}, \ion{Fe}{II}, \ion{Co}{I}, \ion{Ni}{I}, \ion{Ba}{II}, \ion{Sr}{I} and \ion{Sr}{II}\xspace}
\newcommand{\alldetectionsnarrowband}{\ion{H}{$\alpha$}, \ion{H}{$\beta$}, \ion{H}{$\gamma$}, \ion{Li}{I}, \ion{Na}{I}, \ion{K}{I}, \ion{Mg}{I}, \ion{Ca}{II}, \ion{Sr}{I}, \ion{Sr}{II}, and \ion{Mn}{I}\xspace}

\newcommand{\kpvsys}{$K_p-V_{\rm sys}$\xspace}
\newcommand{\V}{\ion{V}{I}\xspace}
\newcommand{\Ti}{\ion{Ti}{I}\xspace}
\newcommand{\allccftemplates}{
\ion{ALL}, \ion{Al}{I}, AlO, \ion{Ba}{I}, \ion{Ba}{II}, 
\ion{Ca}{I}, \ion{Ca}{II}, \ion{Ce}{I}, \ion{Ce}{II}, \ion{C}{I}, 
\ion{Co}{I}, \ion{Co}{II}, \ion{Cr}{I}, \ion{Cr}{II}, \ion{Cs}{I},
\ion{Cu}{I}, \ion{Dy}{I}, \ion{Dy}{II}, \ion{Er}{I}, \ion{Er}{II},
\ion{Eu}{I}, \ion{Eu}{II}, \ion{Fe}{I}, \ion{Fe}{II}, \ion{Ga}{I},
\ion{Gd}{I}, \ion{Gd}{II}, \ion{H2O}{}, \ion{Hf}{I}, \ion{Ho}{I},
\ion{In}{I}, \ion{Ir}{I}, \ion{K}{I}, \ion{La}{I}, \ion{La}{II}, 
\ion{Li}{I}, \ion{Lu}{I}, \ion{Mg}{I}, \ion{Mg}{II}, \ion{Mn}{I}, 
\ion{Mn}{II}, \ion{Mo}{I}, \ion{Na}{I}, \ion{Nb}{I}, \ion{Nd}{I}. \ion{Nd}{II}, \ion{Ni}{I}, \ion{Ni}{II}, \ion{O}{I}, \ion{Os}{I},
\ion{Pb}{I}, \ion{Pd}{I}, \ion{P}{I}, \ion{Pr}{I}, \ion{Pr}{II},
\ion{Rb}{I}, \ion{Rh}{I}, \ion{Ru}{I}, \ion{Sc}{I}, \ion{Sc}{II},
\ion{S}{I}, \ion{Si}{I}, \ion{Si}{II}, \ion{Sm}{I}, \ion{Sm}{II}, \ion{Sr}{I}, \ion{Sr}{II}, \ion{Tb}{I}, \ion{Th}{I}, \ion{Th}{II}, \ion{Ti}{I}, \ion{Ti}{II}, \ion{TiO}{}, \ion{Tm}{I}, \ion{U}{I}, \ion{U}{II}, \ion{V}{I}, \ion{V}{II}, \ion{VO}{}, \ion{W}{I},
\ion{Y}{I}, \ion{Y}{II}, \ion{Zn}{I}, \ion{Zr}{I}, \ion{Zr}{II}
}

\begin{document}

\title{Titanium chemistry of WASP-121 b with ESPRESSO in 4-UT mode}

\titlerunning{}
\authorrunning{}

   \author{
          B.~Prinoth\inst{1,2} 
          \and
          J.~V.~Seidel
          \inst{1,3}\fnmsep\thanks{ESO Fellow}
          \and
          H.~J.~Hoeijmakers\inst{2} 
          \and 
          B.~M.~Morris\inst{4} 
          \and
          M.~Baratella\inst{1} 
          \and
          N.~W.~Borsato\inst{2,5} 
          \and
          Y.~C.~Damasceno\inst{6,7,1} 
          \and
          V.~Parmentier\inst{3} 
          \and 
          D.~Kitzmann\inst{8} 
          \and
          E.~Sedaghati\inst{1} 
          \and
          L.~Pino\inst{9}  
          \and 
          F.~Borsa\inst{10}  
          \and 
          R.~Allart\thanks{SNSF Postdoctoral Fellow}\inst{11} 
          \and
          N.~Santos\inst{6,7}  
          \and
          M.~Steiner\inst{12}  
          \and
          A.~Su{\'a}rez Mascare{\~n}o\inst{13,14}  
          \and 
          H.~Tabernero\inst{15}  
          \and
          M.~R.~Zapatero~Osorio\inst{16}
          }

    \institute{European Southern Observatory, Alonso de C\'ordova 3107, Vitacura, Regi\'on Metropolitana, Chile 
    \and
    Lund Observatory, Division of Astrophysics, Department of Physics, Lund University, Box 118, 221 00 Lund, Sweden 
    \and 
    Laboratoire Lagrange, Observatoire de la Côte d’Azur, CNRS, Université Côte d’Azur, Nice, France 
    \and 
    Space Telescope Science Institute, Baltimore, MD 21218, USA 
    \and 
    School of Mathematical and Physical Sciences, Macquarie University, Sydney, NSW 2109, Australia 
    \and 
    Instituto de Astrofísica e Ci\^encias do Espa\c{c}o, Universidade do Porto, CAUP, Rua das Estrelas, 4150-762 Porto, Portugal 
    \and 
    Departamento de F\'isica e Astronomia, Faculdade de Ci\^encias, Universidade do Porto, Rua do Campo Alegre, 4169-007 Porto, Portugal 
    \and
    University of Bern, Physics Institute, Division of Space Research \& Planetary Sciences, Gesellschaftsstr. 6, 3012, Bern, Switzerland 
    \and 
    INAF — Osservatorio Astrofisico di Arcetri, Florence, Italy 
    \and 
    INAF - Osservatorio Astronomico di Brera, Via E. Bianchi 46, 23807, Merate (LC), Italy 
    \and 
    D\'epartement de Physique, Institut Trottier de Recherche sur les Exoplan\`etes, Universit\'e de Montr\'eal, Montr\'eal, Qu\'ebec, H3T 1J4, Canada 
    \and 
    Observatoire de l’Universit\'e de Gen\`eve, Chemin Pegasi 51, 1290 Versoix, Switzerland 
    \and 
    Instituto de Astrof\'{\i}sica de Canarias, c/ V\'ia L\'actea s/n, 38205  La Laguna, Tenerife, Spain 
    \and 
    Departamento de Astrof\'{\i}sica, Universidad de La Laguna, 38206 La  Laguna, Tenerife, Spain \label{uiac} 
    \and 
    Departamento de F{\'i}sica de la Tierra y Astrof{\'i}sica \& IPARCOS-UCM (Instituto de F\'{i}sica de Part\'{i}culas y del Cosmos de la UCM), Facultad de Ciencias F{\'i}sicas, Universidad Complutense de Madrid, 28040 Madrid, Spain 
    \and
    Centro de Astrobiolog\'\i a, CSIC-INTA, Camino Bajo del Castillo, s/n, E-28692 Villanueva de la Ca\~nada, Madrid, Spain
}
   \date{Received XX XX, 2024; accepted XX XX, 2024}

 
\abstract{Transit spectroscopy usually relies on the integration of one or several transits to achieve the signal-to-noise ratio (S/N) necessary to resolve spectral features. Consequently, high-S/N observations of exoplanet atmospheres, where we can forgo integration, are essential for disentangling the complex chemistry and dynamics beyond global trends. In this study, we combined two partial 4-UT transits of the ultrahot Jupiter WASP-121\,b, observed with the ESPRESSO at the European Southern Observatory's Very Large Telescope in order to revisit its titanium chemistry. Through cross-correlation analysis, we achieved detections of \ion{H}{I}, \ion{Li}{I}, \ion{Na}{I}, \ion{K}{I}, \ion{Mg}{I}, \ion{Ca}{I}, \ion{Ti}{I}, \ion{V}{I}, \ion{Cr}{I}, \ion{Mn}{I}, \ion{Fe}{I}, \ion{Fe}{II}, \ion{Co}{I}, \ion{Ni}{I}, \ion{Ba}{II}, \ion{Sr}{I,} and \ion{Sr}{II}. Additionally, narrow-band spectroscopy allowed us to resolve strong single lines, resulting in significant detections of \ion{H}{$\alpha$}, \ion{H}{$\beta$}, \ion{H}{$\gamma$}, \ion{Li}{I}, \ion{Na}{I}, \ion{K}{I}, \ion{Mg}{I}, \ion{Ca}{II}, \ion{Sr}{I}, \ion{Sr}{II}, and \ion{Mn}{I}. Our most notable finding is the high-significance detection of \ion{Ti}{I} ($\sim 5\sigma$ per spectrum, and $\sim 19\sigma$ stacked in the planetary rest frame). Comparison with atmospheric models reveals that \ion{Ti}{I} is indeed depleted compared to \ion{V}{I}. We also resolve the planetary velocity traces of both \ion{Ti}{I} and \ion{V}{I}, with \ion{Ti}{I} exhibiting a significant blueshift toward the end of the transit. This suggests that \ion{Ti}{I} primarily originates from low-latitude regions within the super-rotating jet observed in WASP-121\,b. Our observations suggest limited mixing between the equatorial jet and the mid-latitudes, in contrast with model predictions from General Circulation Models. We also report the non-detection of TiO, which we attribute to inaccuracies in the line list that could hinder its detection, even if present. Thus, the final determination of the presence of TiO must await space-based observations. We conclude that the 4-UT mode of ESPRESSO is an excellent testbed for achieving high S/N on relatively faint targets, paving the way for future observations with the Extremely Large Telescope.}
\maketitle

\begin{table*}
    \caption{Overview of observations.}%
    \begin{center}
            \begin{tabular}{llllllll}
                    \toprule
                    Phase coverage$^{a}$  & Program ID (PI)              & Date          & $\#$ Spectra$^b$  & t$_{\rm exp}$ [s] & Airmass $^{c}$ & Avg. S/N  & Min./Max. seeing [''] \\
                    \midrule
                    -0.014 -- 0.078       & 1102.C-0744 (GTO)            & 30 Nov 2018    & 29 (19/10)         & 300               & 1.1-1.0-1.1   & 145.9      & 0.55 / 1.05 \\
                    -0.078 -- 0.010      & 111.24J8 (Seidel)            & 23 Sep 2023    & 30 (19/11)         & 300               & 2.3-1.2-1.2   & 148.1      & 0.36 / 0.66 \\
                    \bottomrule
            \end{tabular}
    \end{center}
    \textit{Note:} $^{a}$ Transit from phases $\phi_{\rm start} = -0.045$ to $\phi_{\rm end} = 0.045$. $^{b}$ In parentheses, in-transit and out-of-transit spectra, respectively.  $^{c}$ Airmass at the start and end of the observation, as well as minimum airmass at the highest altitude of the target. 
    \label{tab:observation_log}
\end{table*}

\section{Introduction}

The atmospheres of the hottest known giant planets, ultrahot Jupiters, are subjected to extreme environments due to the proximity to their host stars, leading to the ionisation of atoms, dissociation of molecules, strong global winds, and the presence of metal hydrides, and the potential for thermal inversions by refractory molecules like \ion{TiO}{} and \ion{VO}{} \citep{parmentier_thermal_2018,arcangeli_h-_2018,kitzmann_peculiar_2018,fortney_hot_2021,lothringer_extremely_2018}. Their hot and inflated atmospheres make them unique laboratories for studying atmospheric chemistry and physics under conditions not present in our own Solar System. 

\target \citep[$R_p=$ \SI{1.753 \pm 0.036}{R_{\rm Jup}}, $M_p=$ \SI{1.157 \pm 0.070}{M_{\rm Jup}}, $T_{\rm eq} =$ \SI{2358 \pm 52}{\kelvin}, ][]{delrez_wasp-121_2016,bourrier_hot_2020} is an archetypal ultrahot Jupiter that offers a unique opportunity to probe these extreme atmospheric conditions in detail. It orbits an F-type star with a period of $P \sim$ \SI{1.27}{days}. The planet's orbit is highly misaligned \citep[$\lambda =$ \numpm{87.20}{+0.41}{-0.45}\si{\deg}][]{bourrier_hot_2020} and is close to its Roche limit, suggesting it is close to tidal disruption. The host star, with a magnitude of V = \SI{10.514 \pm 0.006}{mag}, is sufficiently bright to allow high signal-to-noise ratio (S/N) observations, with exposure times short enough to minimise smearing effects \citep{boldt-christmas_optimising_2023}.

Since its discovery in 2016 \citep{delrez_wasp-121_2016}, \target has been extensively studied using various instruments, including the Near InfraRed Spectrograph (NIRSpec) on the JWST \citep{mikal-evans_jwst_2023}, the Wide Field Camera 3 (WFC3) aboard the Hubble Space Telescope (HST) \citep{evans_detection_2016, evans_optical_2018, mikal-evans_diurnal_2022}, the Spitzer Space Telescope \citep{morello_spitzer_2023}, the Ultraviolet/Optical Telescope (UVOT) aboard the Neil Gehrels Swift Observatory \citep{salz_swift_2019}, Gemini Multi-Object Spectrographs (GMOS) \citep{wilson_geminigmos_2021}, the Ultraviolet and Visual Echelle Spectrograph (UVES)  on the ESO VLT \citep{gibson_detection_2020, merritt_inventory_2021, gibson_relative_2022}, the High Accuracy Radial velocity Planet Searcher (HARPS) \citep{hoeijmakers_hot_2020}, the Immersion GRating INfrared Spectrometer (IGRINS) on Gemini South \citep{wardenier_phase-resolving_2024}, ESPRESSO  on the ESO VLT \citep{borsa_atmospheric_2021, maguire_high-resolution_2023, gandhi_spatially_2022, hoeijmakers_mantis_2024}, and even Transiting Exoplanet Survey Satellite (TESS) phase curves \citep{bourrier_hot_2020, daylan_tess_2021}.

However, despite these extensive observations, especially in the optical range, critical questions about the atmospheric properties of \target  remain unresolved. Issues such as the rainout of \ion{TiO}{} \citep{evans_detection_2016, evans_optical_2018, hoeijmakers_hot_2020, bourrier_hot_2020, ouyang_detection_2023}, the presence of ions in the upper atmosphere \citep{sing_hubble_2019, azevedo_silva_detection_2022, maguire_high-resolution_2023}, and atmospheric variability \citep{wilson_geminigmos_2021, maguire_high-resolution_2023, seidel_detection_2023, ouyang_detection_2023, changeat_is_2024} continue to challenge our understanding of this planet. Many of these unresolved issues arise from the limited S/N of the data, which has resulted in non-detections or upper limits for key atmospheric constituents. This limitation complicates theoretical modelling. For example, there remain significant questions regarding the impact of non-local thermal equilibrium (non-LTE) effects on extremely bloated exoplanets like \target \citep{young_searching_2024}.

Ground-based observations have significantly benefited from the high-resolution cross-correlation technique initially introduced by \citet{snellen_orbital_2010} for exoplanet atmospheres; see \citet{birkby2018} for an extensive review. This method has been pivotal in identifying and detecting atoms and molecules in the atmospheres of hot and ultrahot Jupiters \citep{brogi_signature_2012,hoeijmakers_atomic_2018}, enabling detections by averaging the contributions of all available absorption or emission lines within a given wavelength range. Nowadays, species such as \ion{Fe}{I}, \ion{Ti}{I}, and even heavy metals like \ion{Ba}{II} are routinely detected in these atmospheres \citep[e.g.][]{borsato_mantis_2023,azevedo_silva_detection_2022,prinoth_time-resolved_2023}. Complementary single-line narrow-band spectroscopy \citep{wyttenbach_spectrally_2015} resolves strong individual lines like the \ion{Na}{I} D doublet \citep[e.g.][]{langeveld_survey_2022,seidel_detection_2023}, the \ion{Mg}{I} b triplet \citep{prinoth_atlas_2024, damasceno_atmospheric_2024}, hydrogen Balmer lines \citep{fossati_gaps_2023}, the \ion{He}{I} triplet \citep{nortmann_ground-based_2018, allart_spectrally_2018,allart_high-resolution_2019,allart_homogeneous_2023}, and \ion{Ca}{II} \citep{prinoth_atlas_2024}, probing high up in the atmosphere to study chemical abundances and global circulation patterns \citep[e.g.][]{seidel_wind_2020}. Both techniques generally involve and require shifting the data into the planet's rest frame to enhance the S/N of detections.

Recently, there has been a shift towards time-resolved studies using cross-correlation traces and deep single-line spectroscopy \citep[e.g.][]{cauley_time-resolved_2021,pino_gaps_2022,seidel_detection_2023,hoeijmakers_mantis_2024,prinoth_time-resolved_2023,prinoth_atlas_2024}. These studies, which often combine multiple observations or leverage the brightness of certain systems, allow phase-resolved investigations of atmospheric composition and dynamics. This approach marks a significant advancement, moving from global trend analysis to detailed, time-resolved information.

A common requirement for all phase-resolved studies is high S/N data. Achieving this, particularly for fainter host stars beyond the brightest examples, such as KELT-9 and WASP-189, and avoiding the need to average multiple observations, necessitates larger photon-collecting areas. While this challenge could be addressed by waiting for the Extremely Large Telescope (ELT), a more immediate solution is to combine light from several telescopes into a single spectrograph or combining several nights of observations.

The Echelle Spectrograph for Rocky Exoplanets and Stable Spectroscopic Observations (ESPRESSO) offers a solution, with the possibility to simulate a larger telescope. ESPRESSO, an ultra-stabilised, fibre-fed, high-resolution spectrograph located at the ESO VLT on Cerro Paranal, Chile \citep{pepe_espressovlt_2021}, is connected to each of the four unit telescopes (UTs) of the VLT via a separate Coudé train. Typically, ESPRESSO receives light from one UT, which has a primary mirror diameter of 8.2 meters, achieving a maximum resolving power of $R \sim \num{140000}$. However, in its 4-UT mode, ESPRESSO combines light from all four UTs, effectively providing a photon-collecting area equivalent to a 16 m primary mirror. In this mode, the resolving power is $R \sim \num{70000}$ (medium-resolution mode). 


This paper is structured as follows: In Section\,\ref{sec:observation_data_red}, we describe the observations and data reduction process. Section\,\ref{sec:cc-analysis} details our cross-correlation analysis, while in Section\,\ref{sec:narrow_band}, we present our narrow-band analysis. In Section\,\ref{sec:stellar_metallicity}, we revisit the stellar parameters, which are then used in Section\,\ref{sec:titanium_chemistry} to revisit the titanium chemistry. An overarching discussion and conclusions are presented in Section\,\ref{sec:conclusion}.

\section{Observations and data reduction}
\label{sec:observation_data_red}

We used two partial transits of the ultrahot Jupiter \target observed with ESPRESSO as part of the Guaranteed Time Observations by the ESPRESSO Consortium under ESO programme 1102.C-0744 (egress, 30 Nov 2018) and under ESO programme 111.24J8 (PI: Seidel, ingress, 23 Sep 2023). Both partial transits were observed in 4-UT mode at medium resolution ($R \sim \num{70000}$) using 4$\times$2 binning. The primary fibre (fibre A) was placed on the target, while fibre B was positioned on the Fabry-Perot simultaneous reference (for egress) and on sky (for ingress).

Both transit observations have already been analysed, namely by \citet{borsa_atmospheric_2021} for egress and by \citet{seidel_sub} for ingress. A log of the observations is provided in Table\,\ref{tab:observation_log}. The spectra were reduced using the dedicated reduction pipeline (v3.0.0) provided by ESO and the ESPRESSO consortium. From the pipeline data products, we used the two-dimensional (order-by-order) non-blaze-corrected spectra from fibres A and B, and the one-dimensional spectra of the flux-calibrated fibre A to perform the telluric correction.

The reduced spectra were corrected for telluric contamination using \texttt{molecfit} \citep[v4.3.1,][]{smette_molecfit_2015,kausch_molecfit_2015}, as described by \citet{seidel_sub}. Regions with strong \ch{H2O} and \ch{O2} absorption lines around 595, 630, and 647.5 nm were used to fit the model for each exposure individually, selecting small wavelength ranges that contain telluric lines surrounded by a flat continuum without stellar lines. The resulting telluric models were interpolated onto the same wavelength grid as the individual spectral orders and were then divided out to remove telluric effects. 

\section{Cross-correlation analysis}
\label{sec:cc-analysis}

We searched for atomic and molecular absorption by analysing our data with the cross-correlation technique \citep{snellen_orbital_2010} using \texttt{tayph} \citep[see e.g.][]{hoeijmakers_hot_2020,prinoth_titanium_2022,borsato_mantis_2023,tayph_2024_11506199}. The individual, telluric-corrected spectra were Doppler shifted to the rest frame of the star, accounting for the stellar reflex motion due to the orbiting planet, assuming a circular orbit with

\begin{equation}
    v_{\rm Doppler-shift}(\bm{\phi}) = K_{\rm p} \sin{2 \pi \bm{\phi}} = v_{\rm orb} \sin{i} \sin{2 \pi \bm{\phi}},
\end{equation}

where $K_{\rm p}$ is the projected orbital velocity, $v_{\rm orb}$ is the orbital velocity, $i$ is the inclination of the orbit, and $\bm{\phi}$ is the orbital phase; see Table\,\ref{tab:fixed_params} for the parameters. Following \citet{hoeijmakers_high-resolution_2020}, we corrected for outliers in the spectra through an order-by-order sigma clipping algorithm, which calculates the running median absolute deviation over sub-bands of the time series with a width of 40 pixels and rejects 5$\sigma$ outliers. Additionally, we masked any residual telluric contamination caused by deep lines (this is especially noticeable for telluric lines that cause a drop in flux by 50\% and more), particularly around the \ch{O2} band, where deeply saturated lines are not successfully corrected. Manual masking resulted in $\sim $2\% of all pixels being masked out. The spectra were then colour corrected (normalised) using a polynomial of degree 3, accounting for variations in illumination. 
For titanium oxide, we introduced additional masking following \citet{prinoth_titanium_2022} to avoid regions where the line list is known to be imprecise \citep{mckemmish_exomol_2019}. This affects the wavelength ranges up to 460 nm, 507.2–521.6 nm, 568.8–580.6 nm, 590.9–615.4 nm, and 621.0–628.0 nm. For the ESPRESSO 4-UT mode, this corresponds to the spectral orders 0-28, 41-45, 55-57, and 59-65, resulting in 50\% of the spectral pixels being masked out for TiO.

We then cross-correlated the corrected data with 85 templates from \citet{kitzmann_mantis_2023}, whereas we only chose the ones providing significant absorption lines in the wavelength range of ESPRESSO\footnote{In order not to impair the readability of the text, we list here all templates that were considered for the cross-correlation analysis: \allccftemplates, where ALL includes all the species in \citet{kitzmann_mantis_2023} with over $140$ species, including atomic species, individual ions, and H$_2$O, TiO, and CO. }. For the templates, the atmosphere was assumed to be in isothermal and hydrostatic equilibrium at \SI{2500}{\kelvin} for neutrals and molecules and at \SI{4000}{\kelvin} for ions, as they do not absorb significantly at lower temperatures. The templates and models were broadened to the full width at half maximum of \SI{4.28}{\km\per\second}, matching the approximate line-spread function of ESPRESSO in MR mode ($R \sim \num{70000}$).

We corrected for the Doppler shadow of each cross-correlation species using \texttt{StarRotator} \citep[see Hoeijmakers et al., in prep;][]{prinoth_time-resolved_2023,jens_hoeijmakers_2024_13789136}, assuming the average stellar absorption line ---that is, the stellar cross-correlation function--- to be a Gaussian. With a grid size of 75, resulting in 150 $\times$ 150 grid cells in total, the Gaussian in each cell as a function of the radial velocity $v$ is given as

\begin{equation}
    \text{CCF}_i(v) = 1 - A \exp{\left(- \frac{(v_i - v + v_{\rm sys} )^2}{2\sigma^2}\right)},
\end{equation}

where $v_i$ is the velocity of cell $i$, $A$ is the amplitude of the absorption, ranging from 0 (no absorption) to 1 (maximal absorption, saturation), and $\sigma$ is the width of the Gaussian, which is dependent on the instrument. In the stellar rest frame, for a cell at the centre of the stellar surface, this would mean that the velocity is solely given by the systemic velocity $v_{\rm sys}$. Additionally, the velocity of the cell depends on the projected rotational velocity of the host star $v\sin{i}$ and the planet's location with respect to the disc, which in turn depends on the projected orbital obliquity $\lambda$; see \citet{prinoth_atlas_2024} for a full derivation. Additionally, the planet may optimally cover several grid cells to resolve the Doppler shadow sufficiently, requiring more grid cells for smaller the planet-to-star radius ratios $\frac{R_p}{R_\ast}$. We fixed $\frac{R_p}{R_\ast} = \num{0.12355}$ \citep{bourrier_hot_2020}, because of its degeneracy with the amplitude and width of the stellar cross-correlation function. The posterior distributions were retrieved using \texttt{pymultinest}\citep{buchner_x-ray_2014} in a Bayesian framework; see \texttt{StarRotator} \citep{jens_hoeijmakers_2024_13789136} for an application example. After correcting for the Doppler shadow, we removed the stellar lines by dividing the master out-of-transit cross-correlation function. Any residual broadband structures in the spectral direction were then removed using a Gaussian high-pass filter with a width of \SI{100}{km/s}. Lastly, we combined the two data sets, namely the ingress and egress parts, into one two-dimensional cross-correlation function per species.

\begin{figure*}
    \centering
    \includegraphics[width=0.9\linewidth]{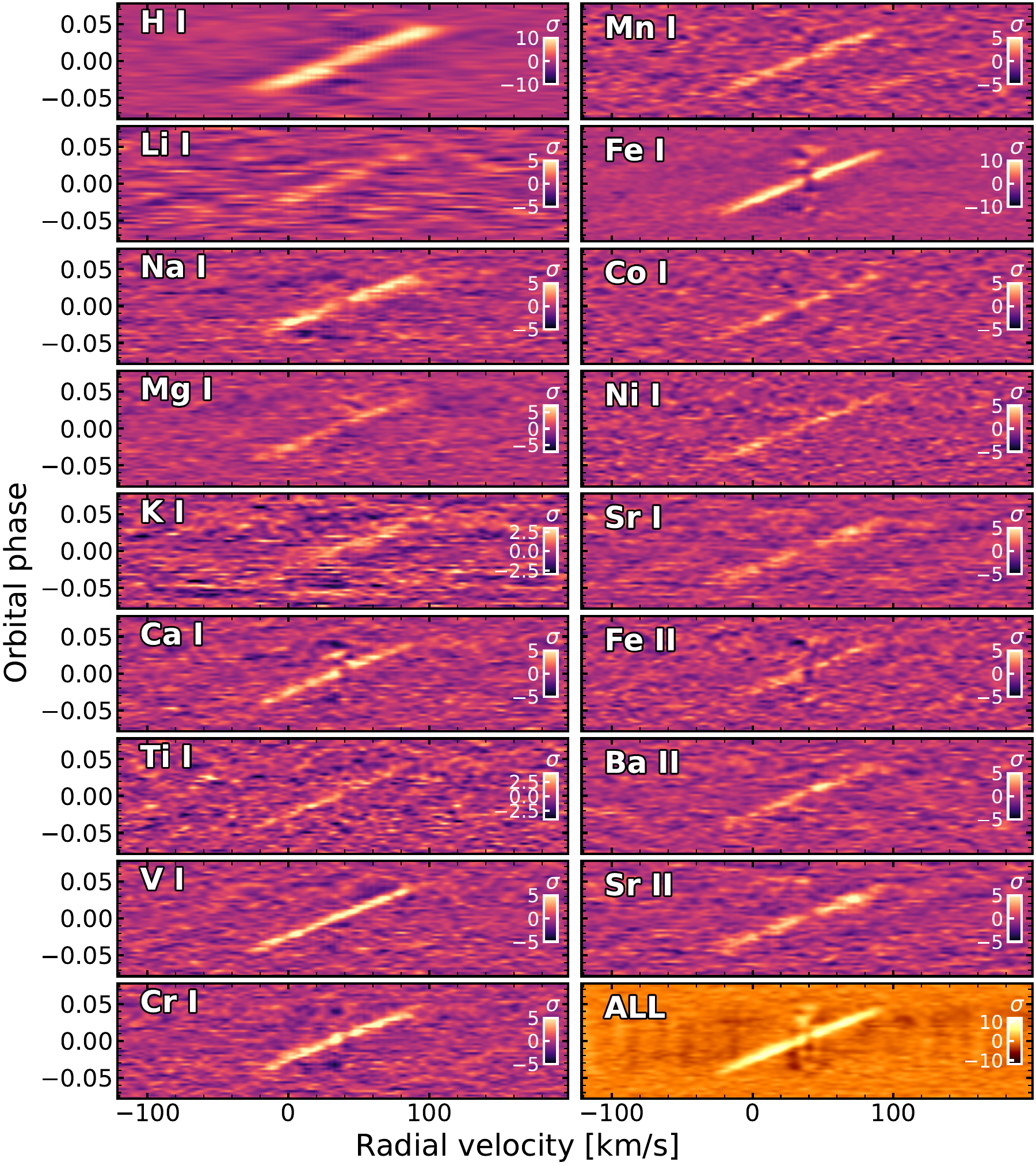}
    \caption{Two-dimensional cross-correlation function of \alldetectionsccf for ingress and egress combined on a common phase grid. The last panel shows the cross-correlation function for the entire atmospheric model ---including all species--- at 2500 K. We highlight the different scales on the colour maps for the deviation from the mean in terms of standard deviations away from the planetary trace, denoted as $\sigma$. We also note that the colour maps have been inverted compared to Fig.\,\ref{fig:ds} to enhance the planetary trace. }
    \label{fig:ccf}
\end{figure*}

\begin{figure*}
    \centering
    \includegraphics[width=\linewidth]{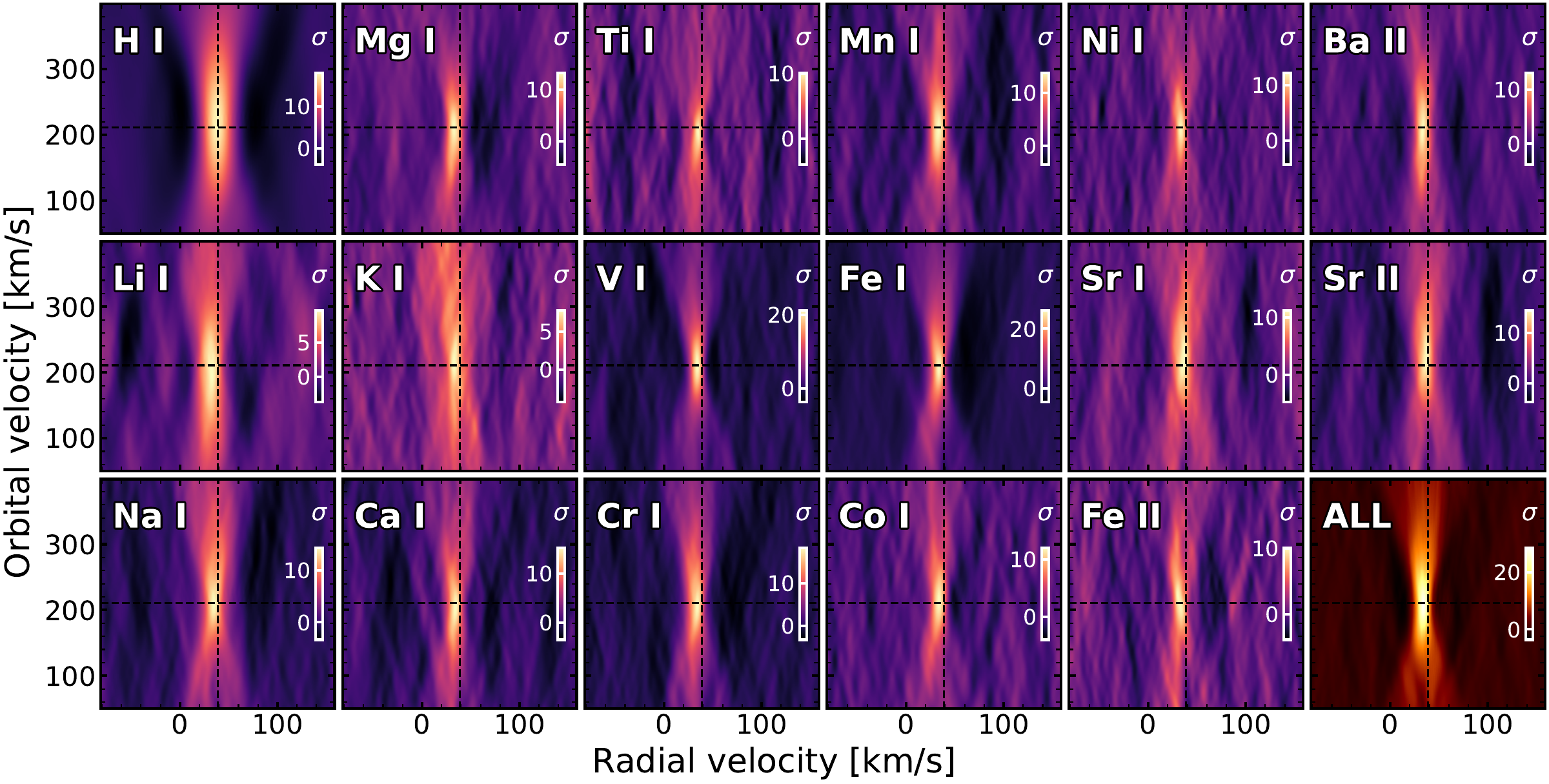}
    \caption{Two-dimensional cross-correlation function of \alldetectionsccf for ingress and egress combined in \kpvsys space. The last panel shows the cross-correlation function for the entire atmospheric model ---including all species--- at 2500 K. We highlight the different scales on the colour maps for the deviation from the mean in terms of standard deviations away from the planetary trace, denoted as $\sigma$.}
    \label{fig:kpvsys}
\end{figure*}

We report detections for \alldetectionsccf in the transmission spectrum of \target through the cross-correlation technique as a time-resolved trace (see Fig.\,\ref{fig:ccf}) and in \kpvsys space (see Fig.\,\ref{fig:kpvsys}). We compute $\sigma$ as the deviation from the mean in terms of standard deviations, where the standard deviation is computed away from the planetary trace. Notably, we detect time-resolved absorption by titanium (see Fig.\,\ref{fig:ccf}), which was previously thought to be non-detectable in the planet's atmosphere due to cold-trapping on the night side; see Section\,\ref{sec:titanium_chemistry} for a full discussion. 

\begin{figure}
    \centering
    \includegraphics[width=0.75\linewidth]{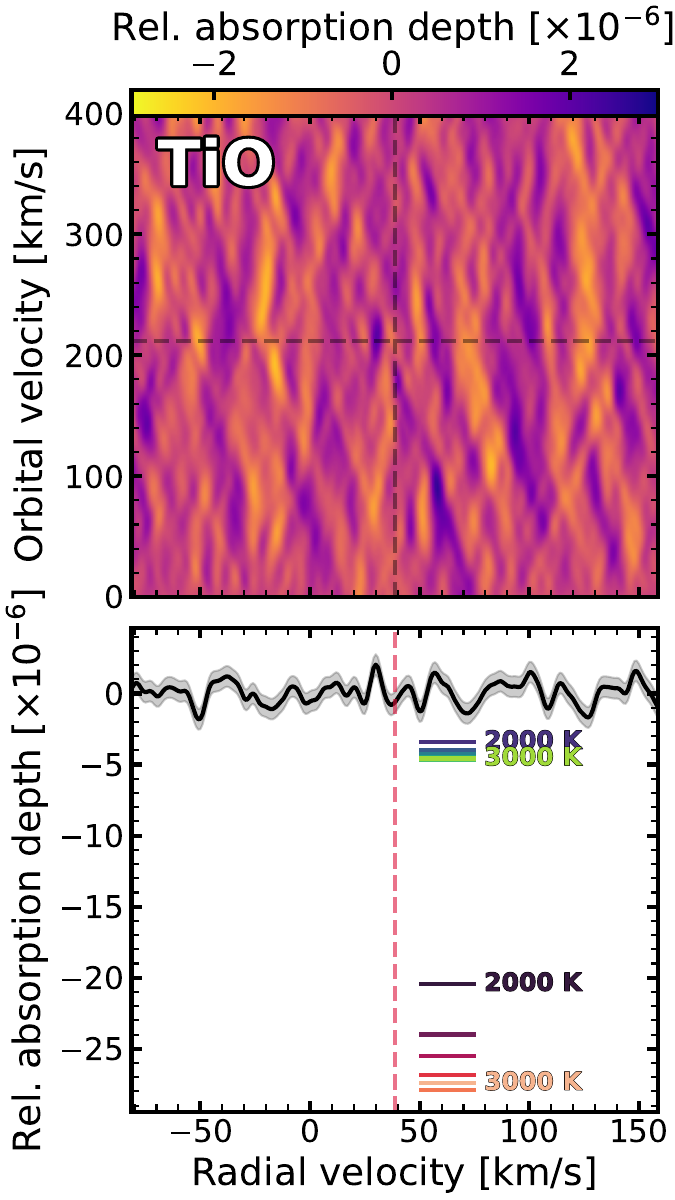}
    \caption{Cross-correlation map of \ion{TiO}{} in \kpvsys space. \textit{Top panel:} Projected orbital velocity (horizontal dashed line) and systemic velocity (vertical dashed line) where the planetary signal is expected. \textit{Bottom panel:} One-dimensional cross-correlation function at the expected orbital velocity. The noise-level is at 10$^{-6}$. The horizontal thermometer lines at the bottom of the panel show the expected absorption depth is predicted by isothermal models for varying temperatures assuming a perfect line list. The temperature step is \SI{200}{\kelvin}. The thermometer lines at the top of the panel are corrected by a factor of 6, empirically suggested by \citet{prinoth_titanium_2022} and \citet{prinoth_time-resolved_2023} to match the observed TiO absorption with the prediction. }
    \label{fig:tio}
\end{figure}

We also searched for TiO absorption using an updated template based on the Toto line list \citep[][states updated on 25 Aug 2021\footnote{ \url{https://www.exomol.com/data/molecules/TiO/48Ti-16O/Toto/}}]{mckemmish_exomol_2019}. The template was computed specifically for \target (see parameters in Table\,\ref{tab:fixed_params}) following \citet{kitzmann_mantis_2023}. In addition to masking telluric residuals, we masked out regions where the TiO line list is known to be imperfect \citep{mckemmish_exomol_2019,prinoth_titanium_2022}. Figure\,\ref{fig:tio} shows the \kpvsys map after combining the ingress and egress datasets, as well as the expected signal strength based on atmospheric models; see Section\,\ref{sec:models}.

\subsection{Aliasing with other atmospheric species}
\begin{figure}
    \centering
    \includegraphics[width=0.9\linewidth]{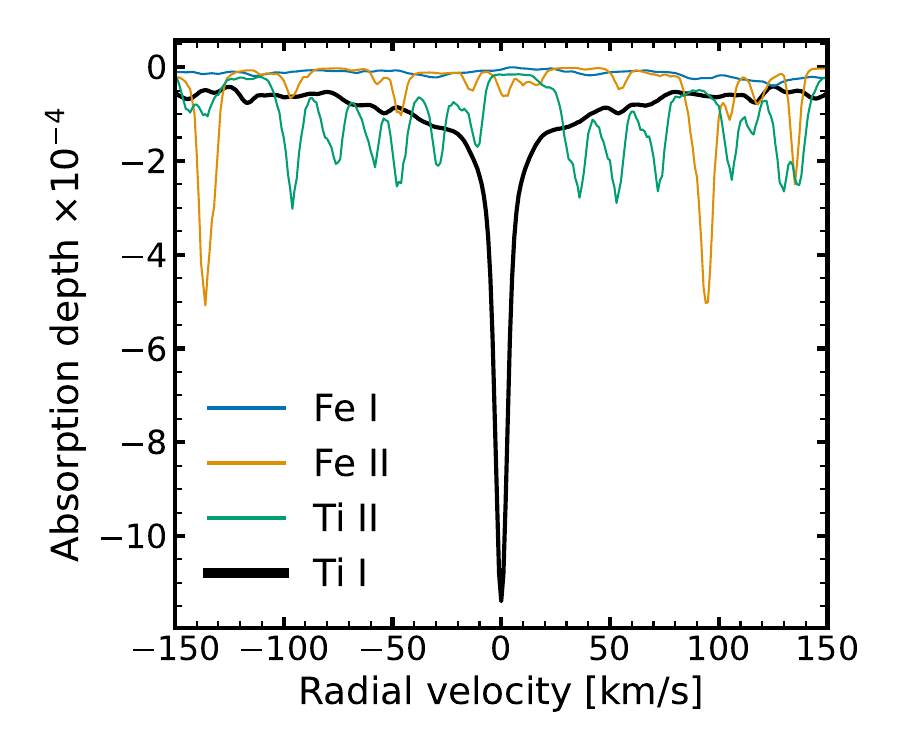}
    \caption{Aliasing of \ion{Ti}{I} at \SI{2500}{\kelvin} with the major absorbing species \ion{Fe}{I} (blue), \ion{Fe}{II} (orange), and \ion{Ti}{II} (green). The black line shows the expected cross-correlation strength for \ion{Ti}{I} at \SI{2500}{\kelvin}. None of the major absorbing species introduce an alias peak at the velocity of the planet.}
    \label{fig:alias}
\end{figure}

Cross-correlation analyses are susceptible to aliasing with other atmospheric species 
\citep{borsato_mantis_2023}. For example, \ion{Fe}{I} produces a secondary peak in the cross-correlation map of \ion{Mg}{I} at a shifted radial velocity \citep{hoeijmakers_hot_2020,prinoth_titanium_2022}, suggesting that the detection of a species could potentially be due to the aliasing of another component. To assess whether the detection of \ion{Ti}{I} is affected by aliasing, we computed the expected alias signals for strong aliasing species at temperatures around \SI{2500}{\kelvin}. At these temperatures, the major absorbing species are \ion{Fe}{I}, \ion{Fe}{II}, and potentially \ion{Ti}{II}. Figure\,\ref{fig:alias} shows the expected aliasing signals computed following the approach used by \citet{borsato_mantis_2023}. Using the calculated templates for cross-correlation, we generate alias profiles by cross-correlating the primary templates used for the detection with templates of other known absorbing species. This produces alias profiles, revealing any secondary peaks at the expected radial velocities where they may occur, which we can then compare with our results. Our efforts demonstrate that none of the major absorbing species introduce an aliasing peak within the radial velocity range of the \ion{Ti}{I} signal, that is, within the planetary rest frame.

\subsection{Model injection}
\label{sec:models}
Similar to \citet{hoeijmakers_mantis_2024} in emission, we compare our detected \Ti and \V signal to the expected signal by a suite of atmospheric models based on the parameters in Table\,\ref{tab:fixed_params}. We used \shone\footnote{\url{https://github.com/bmorris3/shone}} to compute the non-isobaric, isothermal transmission spectrum \citep{de_wit_constraining_2013}, including scattering by hydrogen and helium. The abundances were computed using {\tt FastChem} \citep{stock_fastchem_2018,stock_fastchem_2022,kitzmann_fastchem_2023}. The models were computed at a range of temperatures between \num{2000} and \SI{4000}{\kelvin} for comparison. The model injection follows the same approach as in previous studies \citep[e.g.][]{prinoth_titanium_2022,hoeijmakers_mantis_2024}, where the model is blurred to the spectrograph's resolving power of $R \sim \num{70000}$ using a Gaussian kernel. Additionally, we accounted for broadening due to the planet's rotation. This model is then injected into the raw data through multiplication before any analysis steps. This ensures the same treatment during the cross-correlation analysis as that applied to the real data, in particular masking of telluric regions and exclusion of orders for TiO. To obtain the predicted line depth, the resulting cross-correlation map without model injection is subtracted from the one with the injected model. The expected absorption predicted by the models is shown in Fig.\,\ref{fig:tio} for TiO and Fig. \ref{fig:model} for \Ti and \V. The horizontal thermometer lines show the expected absorption depth predicted by isothermal models for varying temperatures assuming a perfect line list. Implications are discussed in Section\,\ref{sec:titanium_chemistry}.

\subsection{Comparison to previous studies}

The transmission spectrum of \target has been extensively studied using the cross-correlation technique, resulting in detections and confirmations for various atoms and ions, including \ion{H}{I}, \ion{Mg}{I}, \ion{Ca}{I}, \ion{V}{I}, \ion{Cr}{I}, \ion{Fe}{I}, \ion{Ni}{I}, \ion{Fe}{II}, \ion{Ca}{II}, and \ion{K}{I}, \ion{Ba}{II} \citep[e.g.][]{hoeijmakers_hot_2020,gibson_detection_2020,merritt_inventory_2021,borsa_atmospheric_2021,azevedo_silva_detection_2022,maguire_high-resolution_2023}. We confirm all these detections and additionally report detections for \ion{Ti}{I}, \ion{Mn}{I}, \ion{Co}{I} \ion{Sr}{I,} and \ion{Sr}{II}, some of which have previously been claimed tentatively in the \kpvsys maps \citep{hoeijmakers_hot_2020,merritt_inventory_2021}.

We also report the detection of \ion{Li}{I} and \ion{Na}{I}, which were previously detected only using narrow-band spectroscopy. This is due to the limited number of deep lines in the cross-correlation templates at the ESPRESSO wavelengths, favouring narrow-band studies over cross-correlation. For completeness, we add the cross-correlation detections.

\begin{figure}
    \includegraphics[width=\linewidth]{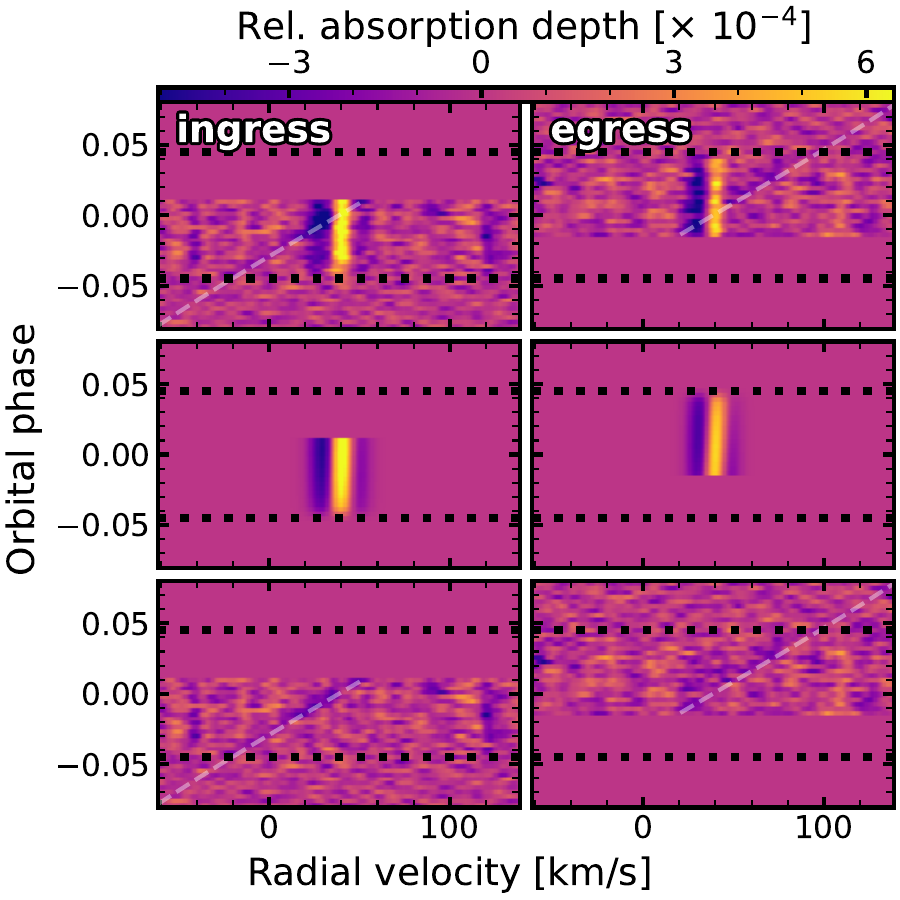}
    \caption{Best-fit Doppler shadow model for \Ti for the ingress (left) and egress (right) epochs. The dashed line indicates the expected location of the planetary trace to guide the eye. The black dotted lines indicate the transit contact times for clarity. We highlight that the colour maps have been inverted compared to Fig.\,\ref{fig:ccf} to reflect planetary absorption as a negative feature. \textit{Upper panels:} \Ti cross-correlation maps prior to correction of the Doppler shadow. \textit{Middle panels:} Retrieved best-fit model for the Doppler shadow. \textit{Bottom panels:} Cross-correlation map after Doppler shadow correction. }
    \label{fig:ds}
\end{figure}

We attribute the capability of detecting \ion{Ti}{I} to the superior photon-collecting power enabled by using ESPRESSO in 4-UT mode compared to a single 1-UT transit, and to improvements in the application of the cross-correlation technique. These include improvements in line lists (not only in terms of accuracy but also in the use of planetary models instead of stellar masks as in \citet{borsa_atmospheric_2021}), the use of two-dimensional orders instead of one-dimensional stitched spectra, and optimisation of the weighting of the cross-correlation templates and the data according to flux changes. The superior S/N achieved in these observations compared to the 1-UT observations in previous studies enables the detection of \ion{Ti}{I} at high significance. \cite{borsa_atmospheric_2021} did not detect \ion{Ti}{I} in the 4-UT egress data set, which we attribute to differences in the application of the cross-correlation technique and the removal of the Doppler shadow. \citet{borsa_atmospheric_2021} applied the cross-correlation technique to one-dimensional stitched spectra produced by the ESPRESSO pipeline, which are already corrected for the blaze function. To properly account for the larger uncertainties at the order edges—due to lower flux—, the cross-correlation function (see Eq.\,(1) in \citet{borsa_atmospheric_2021}) would need to include a division by the uncertainties when applied to 1D spectra (s1d). We apply the same definition of the cross-correlation function, but use the two-dimensional spectra without blaze correction, which explicitly accounts for the lower flux at the order edges. We also correct for the Doppler shadow contamination, whereas \citet{borsa_atmospheric_2021} masked the affected regions, roughly around mid-transit, as seen in Fig.\,\ref{fig:ds}. This affects about half of the in-transit exposures for the egress dataset, removing a significant portion of the planetary signal and thereby impeding the detection of the \Ti signal in the remaining data.



\section{Narrow-band spectroscopy}
\label{sec:narrow_band}

\begin{figure*}
   \sidecaption
   \includegraphics[width=0.693\textwidth]{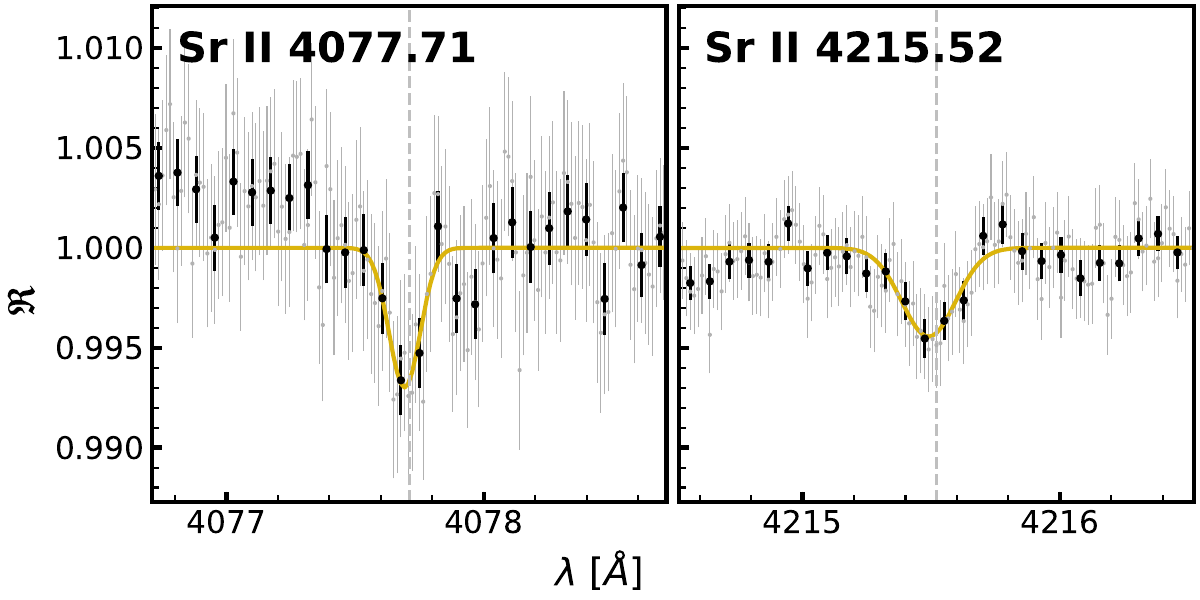}
    \caption{\ion{Sr}{II} shown for the ingress and egress dataset combined. The original data are shown in grey, binned by 5 for better visibility in black, and the Gaussian fit is shown as a solid golden  line. \vspace{4cm}
}
     \label{fig:SrII_narrowband}
\end{figure*}

\begin{figure*}
   \centering
   \includegraphics[width=\textwidth]{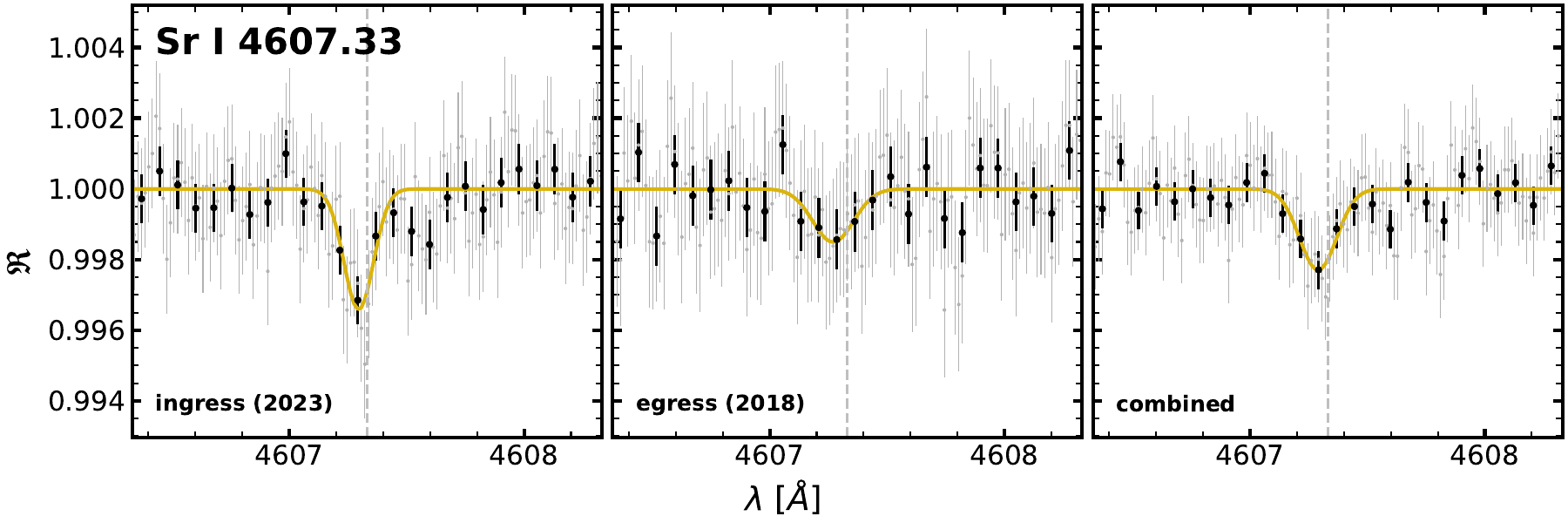}
    \caption{Same as Fig.\,\ref{fig:SrII_narrowband} but for \ion{Sr}{I} for the ingress data set only.}
     \label{fig:SrI_narrowband}
\end{figure*}

\begin{figure*}
   \centering
   \includegraphics[width=\textwidth]{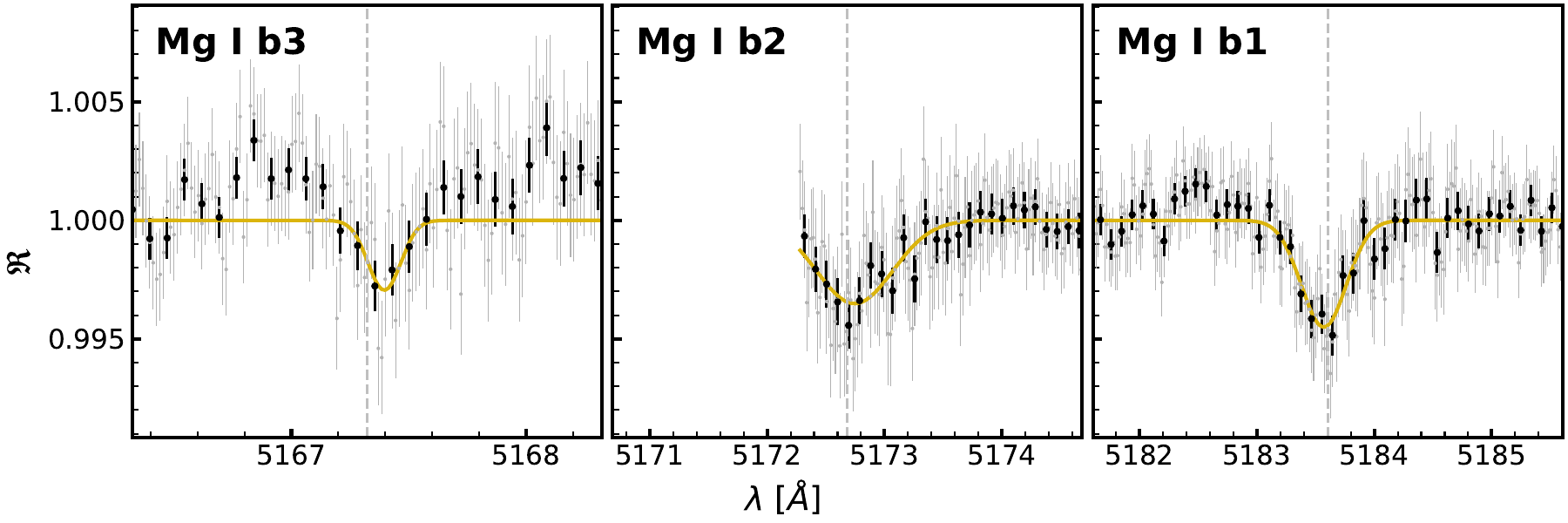}
    \caption{Same as Fig.\,\ref{fig:SrII_narrowband} but for \ion{Mg}{I} b triplet components with the $b_3$, $b_2$, and $b_1$ lines from left to right for the ingress data set only. The $b_2$ line lies at the edge of order $43$ explaining the lack of data on the left.}
         \label{fig:MgItriplet_narrowband}
   \end{figure*}

In addition to our search for elements in cross-correlation, we also directly resolved spectral lines. We applied the same telluric contamination correction as for the cross-correlation analysis and corrected for the Rossiter-McLaughlin (RM) effect in planetary lines with a stellar equivalent via {\tt StarRotator} \citep[see Hoeijmakers et al., in prep;][]{prinoth_time-resolved_2023,jens_hoeijmakers_2024_13789136} with a grid size of 300x300. The telluric contamination was corrected in the observer's rest frame, and the RM effect was corrected in the stellar rest frame, together with the stellar spectrum itself. All further steps in the analysis can be found in \citet{seidel_sub}, and these follow the analysis outlined in \citet{SeidelPrinoth2023}.

A wide range of elements were detected in \citet{borsa_atmospheric_2021} from the egress dataset of the 4-UT transit. In this work, we do not aim to reproduce their results but instead expand on them. While we provide detection levels in Table \ref{tab:narrowdetections} for all species found in \citet{borsa_atmospheric_2021}, we only discuss the additional detections from the 2023 data presented in this work. An additional re-analysis of the egress dataset ---to provide higher detection levels by combining the datasets--- was not necessary in most cases. The generally improved seeing and S/N in the 2023 data, covering the first half of the transit, are most likely the source of the improved data quality in the 2023 observations. Our searches for resolved spectral lines of \ion{Fe}{I}, \ion{Fe}{II}, \ion{Ra}{I}, \ion{Ra}{II}, and \ion{Ba}{I} were unsuccessful, primarily due to the blending and aliasing of the lines with other elements prominent in the studied wavelength ranges. 

\subsection{Comparison to detections with the egress dataset}

The analysis presented here confirms all detections in narrow band from the egress partial transit as published in \citet{borsa_atmospheric_2021}. The detection contrasts, line widths (full width at half maximum), and line centre offsets are reported in Table \ref{tab:narrowdetections}. The datasets used for each detection are indicated in the final column. Consistent detection levels were found for all species except for \ion{K}{I}. In the ingress dataset presented here, due to differences in relative velocity shifts compared to the egress dataset, the \ion{K}{I} line at $7698.96~\AA$ falls directly between orders 81 and 82. Consequently, we did not use the two-dimensional order-by-order data to recover the detection but instead used the full one-dimensional stitched
  spectrum. Nonetheless, the wavelength region is likely affected by edge effects, and our \ion{K}{I} detection is significantly lower in contrast compared to the egress dataset. Additionally, the \ion{K}{I} detection recovered from the ingress dataset is much broader than anticipated, which calls the tentative detection into question. However, given the clarity of the detection in \citet{borsa_atmospheric_2021} from the egress data, we conclude that the uncertainty of our detection is not due to the planetary atmosphere but rather to unfortunate instrumental circumstances.

For elements probing the upper atmosphere, notably the \ion{Ca}{II} H and K lines, the first two lines of the Balmer series, and the \ion{Na}{I} doublet, we can confirm the detection contrast reported by \citet{borsa_atmospheric_2021}, though with different line centre offsets. The line centre offset usually indicates movement of the atmosphere due to planetary rotation and/or winds. However, we observe generally smaller offsets than those reported for these species in \citet{borsa_atmospheric_2021}. We confirm the detection of the Balmer series as presented in \citet{borsa_atmospheric_2021}, where \ion{H}{$\alpha$} and \ion{H}{$\beta$} were detected. However, even when combining the two datasets, we were unable to detect \ion{H}{$\gamma$}, which is obscured by stellar noise. Due to the high noise level, we do not provide an upper limit, as any estimate would not be conducive to further conclusions.

\subsection{Additional detections of alkaline Earth metals}

In addition to the reported detections in \citet{borsa_atmospheric_2021}, we also confidently detect strontium both as \ion{Sr}{II} and \ion{Sr}{I} , which are shown in Figs.\,\ref{fig:SrII_narrowband} and \ref{fig:SrI_narrowband} (the first in combination with the egress dataset). The continuum for the \ion{Sr}{II} line at \SI{4077.71}{\AA} deviates slightly from 1 due to residual systematic uncertainties caused by internal reflections within the Coudé train still visible in the continuum \citep[e.g.][]{allart_wasp-127b_2020}. We optimised our extraction for the Sr II line centre, which means the systematic uncertainties are not fully corrected away from this line, leading to deviations from the baseline. We confirm the cross-correlation detection of \ion{Ba}{II} from \citet{azevedo_silva_detection_2022} and provide tentative narrow-band detections in Figure \ref{fig:BaII_narrowband}. Considering that these detections together with the established detections of \ion{Ca}{II} and \ion{Mg}{I} cover most of group II of the periodic table, that is, the alkaline Earth metals, we also investigated \ion{Ba}{I} and \ion{Ra}{I} with their first ionisation stages. Of these four, only \ion{Ra}{II} has noticeable absorption lines in the visible wavelength range covered by ESPRESSO, at $3814.40~\AA$ and $4682.24~\AA$. Neither of these lines was measurable beyond the continuum, lending further strength to the detection of \ion{Sr}{II} at $4215.52~\AA,$ which aliases with a minor \ion{Ra}{II} line.


\citet{borsa_atmospheric_2021} obtained robust detections of \ion{Mg}{I} via the single line at $4571.1~\AA$ as well as the \ion{Mg}{I} b triplet combined in wavelength space. The \ion{Mg}{I} b triplet stems from the atomic levels $3^3 P_{0,1,2}$ and $4^3 D$, with the $b_1$ line at $5183.60 \AA$ with $J_l =2$, the $b_2$ line at $5172.68 \AA$ with $J_l =1$, and the $b_3$ line at $5167.32 \AA$ with $J_l =0$. We can resolve this triplet with the ingress data, which are shown in Fig.\,\ref{fig:MgItriplet_narrowband}.

\begin{figure}[h!]
    \includegraphics[width=0.9\linewidth]{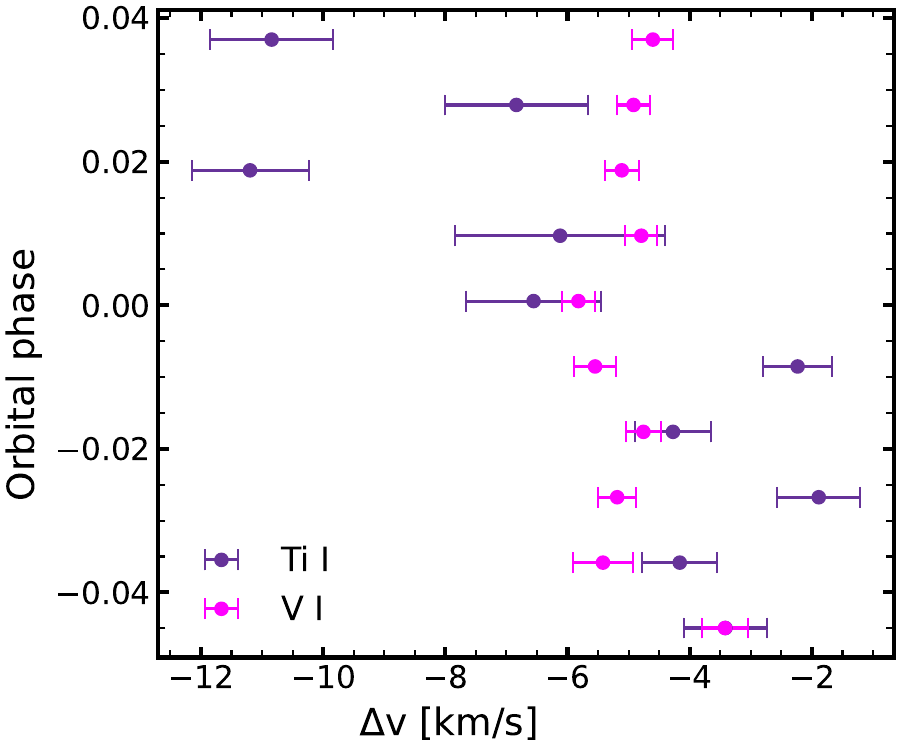}
    \includegraphics[width=0.9\linewidth]{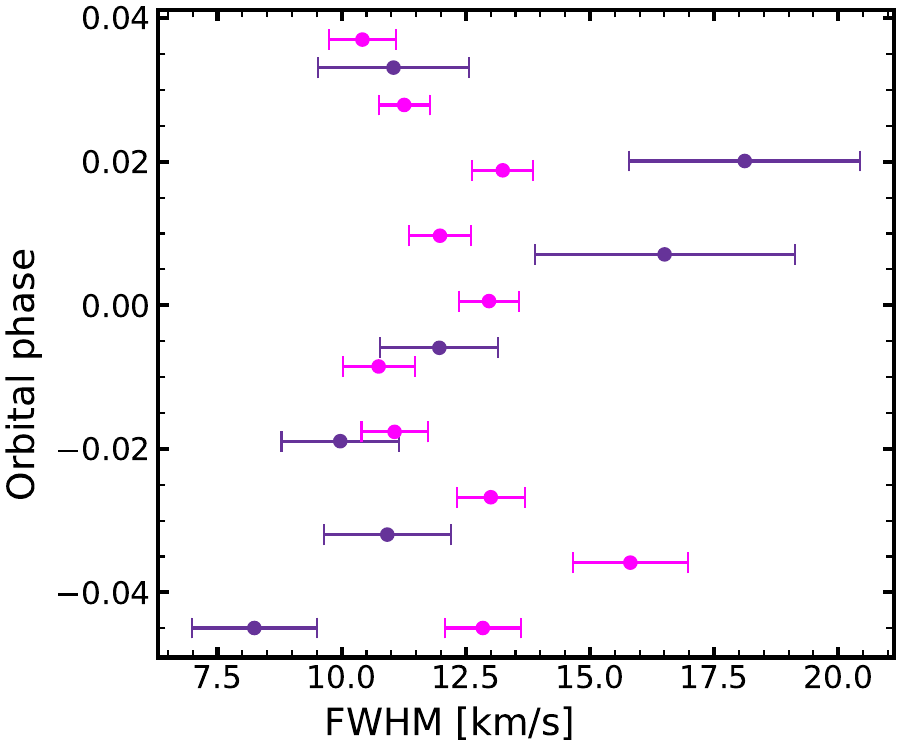}
    \caption{Binned velocity traces (each of three exposures) relative to the systemic velocity (top) and FWHM (bottom)  for \ion{Ti}{I} and \ion{V}{I}. The uncertainties are estimated through JAX; see the explanation in Appendix \ref{app:true_limbs}. Towards the end of the transit, the velocity shift and FWHM for \Ti are larger than those for \V. The velocity shift difference is significantly larger, while for the FWHM, the uncertainties are relatively large, particularly
in \Ti.}
    \label{fig:ehrenreich_effect}
\end{figure}

\section{Stellar parameters}
\label{sec:stellar_metallicity}
To refine the stellar parameters and chemical abundances, we performed a detailed spectroscopic analysis of both the master ingress and egress spectra. The master spectra were created by averaging all spectra of the time-series after correction for tellurics, sky emission, and cosmic rays in the rest frame of the star. 

We derived the effective temperature ($T_{\mathrm{eff}}$), surface gravity ($\log g$), micro-turbulence velocity parameter (V$_{\mathrm{mic}}$), and iron abundance ([Fe/H]\footnote{[Fe/H] = A(Fe)${\star}$ - A(Fe)${\sun}$, where we adopt the solar abundance scale from \cite{asplund_chemical_2021}}) using the classical equivalent width (EW) method. Briefly, we determined the stellar parameters by simultaneously imposing three spectroscopic equilibria: the excitation equilibrium to derive $T{\mathrm{eff}}$, the ionisation equilibrium to derive $\log g$, and V$_{\rm{mic}}$ by ensuring that weak and strong iron lines result in the same abundances. Finally, iron and titanium abundances were inferred using the line list from \cite{baratella_gaps_2020}, which includes a carefully selected set of well-isolated lines optimised for the analysis of planet-host stars. The EWs were measured using the ARES v2 software \citep{sousa_ares_2015}, which fits a Gaussian profile to the lines and excludes those with uncertainties of greater than 10\%. We then used the \textit{qoyllur-quipu} (q$^2$) code\footnote{\url{https://github.com/astroChasqui/q2}} \citep{ramirez_solar_2014} and adopted 1D-LTE Kurucz model atmospheres.

The stellar parameters derived from the ingress and egress spectra are in excellent agreement, with $T_{\mathrm{eff}}$ values of 6790$\pm$96\,K and 6752$\pm$99\,K, $\log g$ values of 4.36$\pm$0.20 and 4.32$\pm$0.19, and V$_{\rm{mic}}$ values of 1.70$\pm$0.11 km s$^{-1}$ and 1.72$\pm$0.10 km s$^{-1}$, respectively. The iron abundances for the two spectra are [\ion{Fe}{I}/H] = 0.33$\pm$0.05 and 0.31$\pm$0.06 (the uncertainties reflect the scatter in the EW measurements). We also derived titanium abundances, which amount to [\ion{Ti}{I}/H] = 0.29$\pm$0.03 and 0.26$\pm$0.03. These super-solar abundances are expected in light of the giant planet--metallicity correlation \citep{adibekyan_heavy_2019, biazzo_gaps_2022}, which predicts a higher frequency of giant planets around 
stars of super-solar metallicity. Using q$^2$, we also derived the stellar radius and mass based on this new set of spectroscopic parameters, obtaining R$_{\star}$ = 1.40$\pm$0.05 R$_{\sun}$ and M${\star}$ = 1.42$\pm$0.03 M$_{\sun}$, which agree reasonably well with values reported in the TESS Input Catalogue \citep{stassun_revised_2019, paegert2021}. \\

We note that the stellar radius of R$_{\star} = 1.40 \pm 0.05$\,R$_\sun$ is smaller than the one reported in \citet{delrez_wasp-121_2016} ($R_{\star} = 1.458 \pm 0.030$\,R$_\sun$), which was also used by \citet{bourrier_hot_2020}. This leads to a decrease in the derived semi-major axis and orbital velocity (see Table\,\ref{tab:fixed_params}) compared to \citet{bourrier_hot_2020}, who obtained $a = $\,\numpm{0.02596}{+0.00043}{-0.00063} AU and $v_{\rm orb} =$ \numpm{221.52}{+0.37}{-0.54} \si{\km\per\second} (derived from the scaled semi-major axis in \citet{bourrier_hot_2020} and the stellar radius in \citet{delrez_wasp-121_2016}). However, the uncertainty in the stellar radius reported in the present work introduces a larger uncertainty in the orbital velocity, placing the derived value for the parameters in \citet{bourrier_hot_2020} within two standard deviations. For consistency, we have used our derived orbital velocity throughout this manuscript.

\section{Titanium chemistry revisited}
\label{sec:titanium_chemistry}

\begin{figure*}
    \centering
    \sidecaption
    \includegraphics[width=0.6\linewidth]{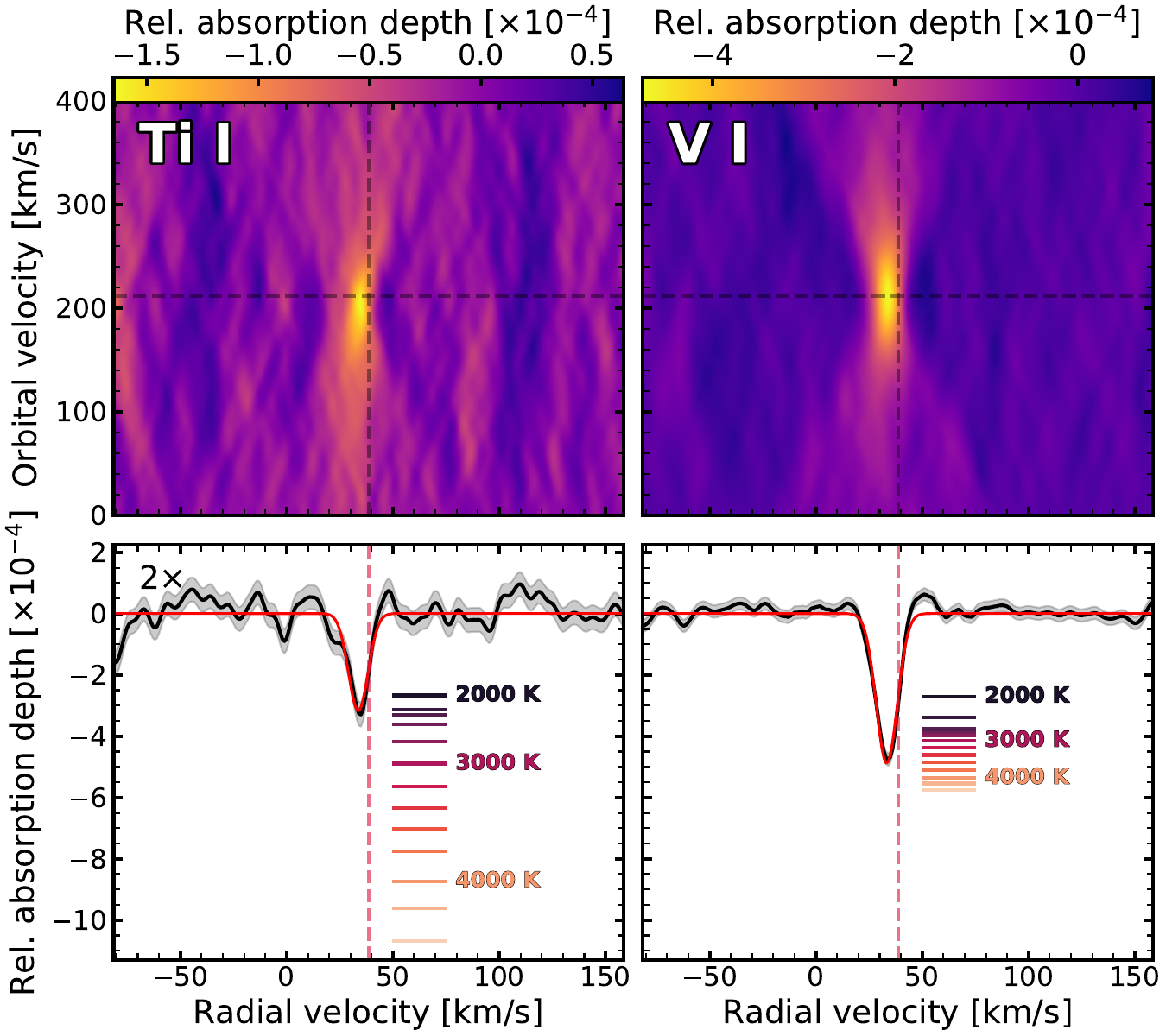}
    \caption{Cross-correlation results in \kpvsys space for \Ti (left) and \V (right). \textit{Top panels:} Two-dimensional cross-correlation map. The dashed lines indicate the expected location of the signal compared to a purely Keplerian orbit. \textit{Bottom panels:} One-dimensional cross-correlation function at the orbital velocity of the strongest absorption from the upper panel. The \Ti panel is multiplied by two to ensure visibility. The horizontal lines indicate the expected absorption depth at the indicated model temperature. The increase between lines is \SI{100}{\kelvin}. \V requires a temperature of $\sim$\SI{3900}{\kelvin}, while for \Ti, the thermometer suggests $\sim$\SI{2900}{\kelvin}. Together with Fig.\,\ref{fig:ehrenreich_effect}, this suggests that \Ti is significantly depleted compared to \V.\vspace{4cm}}
    \label{fig:model}
\end{figure*}

\begin{table*}[]
    \centering
    \caption{Fixed planetary, stellar, orbital, and atmospheric parameters.}
     \begin{tabular}{llll}
     \toprule
        Parameter & Value & Unit & Reference \\
        \midrule
        Planetary radius $R_p$ & \num{1.753 \pm 0.036} & \si{R_{\rm Jup}} & \cite{bourrier_hot_2020}\\
        Planetary mass   $M_p$ & \num{1.157 \pm 0.070} & \si{M_{\rm Jup}} & \cite{bourrier_hot_2020}\\
        Planetary surface gravity   $\log g_p$ & \num{3.0 \pm 1.8} & \si{\cm\per\s\squared} & derived \\
        Orbital period $P$ & \numpm{1.27492504}{+0.00000015}{-0.00000014} & days & \cite{bourrier_hot_2020} \\
        Orbital inclination $i$ & \num{88.49 \pm 0.16} & \si{\deg} & \cite{bourrier_hot_2020} \\
        Scaled semi-major axis $a / R_\star$ & \numpm{3.8131}{+0.0075}{-0.0060} & & \cite{bourrier_hot_2020} \\
        Semi-major axis $a$ & \num{0.024826 \pm 0.00089} & au & derived \\
        Orbital velocity $v_{\rm orb}$ & \num{211.8 \pm 7.6} & \si{\km\per\s} & derived \\
        Systemic velocity $v_{\rm sys}$ & \num{38.64 \pm 0.06} & \si{\km\per\s} & \citet{seidel_sub} \\ 
        Radial velocity semi-amplitude $K$ & \numpm{0.177}{+0.0085}{-0.0081} & \si{\km\per\s} & \cite{bourrier_hot_2020} \\
        Stellar radius $R_\star$ & \num{1.40 \pm 0.05} & \si{R_\odot} &  This work\\
        Stellar mass $M_\star$ & \num{ 1.42 \pm 0.03} & \si{M_\odot} &  This work\\
    \bottomrule
    \end{tabular}
    \label{tab:fixed_params}\\ \vspace{1em}
    \textit{Note:} Most important orbital, planetary, and stellar parameters used in this study.
\end{table*}

In this study, we detect \ion{Ti}{I} at a high significance of $\sim5\sigma$ per spectrum (along the resolved trace); see Fig.\,\ref{fig:ccf}. Stacked in the planetary rest frame, this increases to $19\sigma$; see Fig.\,\ref{fig:model}\footnote{$\sigma$ for the rest frame is based on the standard deviation of the best-fit Gaussian in Fig.\ref{fig:model} using \texttt{lmfit} \citep{newville_2015_11813}.}. In the following, we discuss the implications of this finding for the chemistry and dynamics on \target  and compare to previous studies. \\

The titanium chemistry of \target has been a long-standing topic of debate in the literature since the observation of complex variability at optical wavelengths and a suspected detection of vanadium oxide \citep{evans_optical_2018}. Previous studies at high spectral resolution reported marginally tentative detections or non-detections 
of both \Ti and TiO in emission \citep{hoeijmakers_mantis_2024} and transmission \citep{gibson_detection_2020,merritt_inventory_2021,hoeijmakers_hot_2020,gibson_relative_2022,borsa_atmospheric_2021}. Retrieval analyses of the transmission spectrum also reveal that the \ion{Ti}{I} abundance relative to the star is depleted, 
while \ion{V}{I} remains consistent with the stellar abundance \citep[e.g.][]{maguire_high-resolution_2023,gandhi_retrieval_2023}.

Similar behaviour has recently been observed for WASP-76\,b \citep{pelletier_vanadium_2023}, where the \ion{V}{I} abundance is also consistent with the stellar abundance, while the \ion{Ti}{I} abundance is relatively depleted. \citet{cont_atmospheric_2022} reported the detection of \Ti and \V on the day side of WASP-33\,b, a planet significantly hotter \citep[$T_{\rm eq} \sim \SI{2600}{\kelvin}$;][]{hooton_spi-ops_2022} than both \target and WASP-76\,b. Similarly, \citet{prinoth_time-resolved_2023} presented strong absorption signatures for both \Ti and \V in the transmission spectrum of WASP-189\,b \citep[$T_{\rm eq} \sim \SI{2600}{\kelvin}$;][]{anderson_wasp-189b_2018}. This suggests that both WASP-33\,b and WASP-189\,b are hot enough to avoid significant depletion of \Ti through rainout or condensation, but may instead be subject to ionisation on the day side, as supported by the detection of \ion{Ti}{II} in both planets, and discussed in \citet{prinoth_time-resolved_2023} for WASP-189\,b. Conversely, \citet{scandariato_pepsi_2023} and \citet{guo_detection_2024} reported the detection of \Ti but not \V on the day side of MASCARA-1\,b \citep[$T_{\rm eq} \sim \SI{2800}{\kelvin}$;][]{chakrabarty_precise_2019}, despite its similar equilibrium temperature to WASP-33\,b and WASP-189\,b. \citet{scandariato_pepsi_2023} did not search for \ion{V}{II}, while  \citet{guo_detection_2024}
did not detect \ion{V}{II} due to limited wavelength coverage in the blue caused by the exclusion of wavelengths shorter than 540\,nm when analysing the VIS arm of CARMENES. This means the possibility remains that \V may be significantly ionised instead.
For the cooler planet HD\,149026\,b \citep[$T_{\rm eq} \sim \SI{1600}{\kelvin}$; $T_{\rm day} \sim \SI{2100}{\kelvin}$,][]{southworth_homogeneous_2010,ishizuka_neutral_2021}, a hot Saturn, \citet{ishizuka_neutral_2021} reported a tentative detection of \Ti (< 5$\sigma$); however, \citet{biassoni_high-resolution_2024} could not confirm this detection.

By carrying out model injection (Fig.\,\ref{fig:model}), we determined that the present observations for \target show a behaviour of depletion similar to that seen for planets with similar temperatures ---in particular WASP-76\,b--- , where the \V detection is consistent with a model at $\sim$\SI{3800}{\kelvin}, but \Ti is weaker than expected at this temperature. For \Ti, the observed absorption depth can be explained by a temperature of only $\sim$\SI{2600}{\kelvin} , indicating that \ion{Ti}{I} has a lower abundance than \ion{V}{I} when integrated over the entire limb. \\

These and previous observations necessitate the presence of a physical mechanism that depletes \ion{Ti}-bearing species both at high altitudes across the thermal inversion on the day side and at the terminator region, assuming that the bulk of the planet is not, in reality, depleted in Ti compared to V. These findings also suggest the day-night-side cold-trap phenomenon may be at play \citep{parmentier_3d_2013}. \citet{mikal-evans_diurnal_2022,hoeijmakers_mantis_2024} hypothesised that this depletion is expected because perovskite, \ch{CaTiO3}, condenses between $\sim\num{1600}$ and $\SI{2000}{\kelvin}$ for pressures of between 0.1 bar and 10 bar \citep[see Table 2 in][]{lodders_titanium_2002}, in line with the observed night-side temperature. However, after condensation on the night side, the cloud particles remain coupled to the atmospheric flow and are eventually transported back to the day side. If the cloud particles undergo rainout on the night side, they settle deeper in the atmosphere. Upon recirculation to the day side, they then reside at altitudes below the thermal inversion that are too low to be observed in either day-side emission spectra or transmission spectra \citep{parmentier_3d_2013}. Our observations of titanium in the transmission spectrum are, at first glance, in tension with this picture.

Upon investigation of the velocity shifts during transit \citep[see Fig.\ref{fig:ehrenreich_effect}, similar to Fig. 6 in][]{borsa_atmospheric_2021}, it becomes evident that \Ti exhibits increasing blueshifts towards the end of the transit, while remaining consistent with the rest frame of the planet during the first half. In contrast, \V shows a lower velocity shift and a smaller increase in velocity throughout the transit. To explain the increased velocity shift of \Ti towards the end of the transit,  \Ti must be contained at the equator at the evening, or trailing, terminator. This would allow it to reach larger velocities than if it were distributed over all latitudes, because the planet rotation is larger at the equator and the equator regions likely exhibit a super-rotating flow \citep{showman_equatorial_2011}.

This interpretation is strengthened by the broadened FWHM of \Ti; see Fig.\,\ref{fig:ehrenreich_effect}. A species concentrated in the equatorial jet would exhibit opposing velocity components from the leading (redshifted) and trailing (blueshifted) limbs, producing a broader signal. In contrast, \V, which appears more uniformly distributed, would result in a narrower velocity profile dominated by material near zero velocity: Although absorption from \V from equatorial latitudes could produce pronounced line wings, the FWHM ultimately depends on the ratio of globally distributed to jet-localised material. For a jet-localised species, however, the FWHM is expected to be broad, or may even be double peaked \citep{Nortmann2024}.

The increasing blueshift of \Ti, along with its broad FWHM, strongly suggests a preferential equatorial distribution, while the smaller velocity shifts of  \V\  and its narrower FWHM imply a more uniform terminator distribution. However, the uncertainties on the FWHM of \Ti, due to its weaker signal, remain significant, and the apparent differences for the FWHM between the two species may not be statistically robust.

The apparently small signal of Ti could also be explained if Ti were contained in atmospheric regions with significantly smaller scale heights than \V (e.g. due to a lower temperature). This would be the case if \Ti and \V were to primarily occur on opposing (leading versus trailing) limbs with a strong temperature difference. To exclude this scenario, we note that \Ti remains visible towards the end of the transit, while expressing a significant blueshift, implying its presence on the trailing limb that is rotating towards the observer, and also highlight the fact that there is no known mechanism or evidence that \V is depleted from those regions. We also averaged the true limb exposures (first and last three exposures in ingress and egress respectively); see Fig.\,\ref{fig:true_limbs}, and find that \Ti is indeed present at both limbs, which we can resolve thanks to the high S/N of these observations; though we note that the egress observation generally has a lower S/N. 

We propose that the weak signature of \Ti and its strong blueshift towards the end of the transit event can instead be explained by a combination of two factors: a significant depletion of \Ti compared to \V, and absorption over a smaller fraction of the terminator compared to \V. The observed absorption of \Ti originates preferentially from the equatorial regions, while \V is more uniformly distributed over the terminator region. However, \V also appears blueshifted with respect to the systemic velocity, which may be indicative of a day-to-night-side flow. A detailed investigation of this potential flow is beyond the scope of this paper.

The confinement of Ti material within equatorial latitudes, while \V is homogeneously distributed, implies that mixing between the equatorial jet and the mid-latitudes is limited. In a situation where the mid-latitudes are depleted of Ti, it would likely not be replenished by the material within the equatorial jet. This equatorial jet versus mid-latitude behaviour is seen in most global circulation models \citep[see Fig. 13 of the review][]{showman2021}. However, when a condensable molecule is added to these models, the equatorial zone tends to be depleted in tracer, which would go in the opposite direction to what the current observations imply. Our observations therefore call for further work into understanding atmospheric mixing in these atmospheres.

As in previous studies, model predictions appear to strongly rule out the presence of TiO \citep[see e.g.][]{hoeijmakers_mantis_2024}, which would be in tension with the clear detection of \Ti. Detections of TiO are naturally hindered by inaccuracies in the TiO line list \citep{hoeijmakers_search_2015}, and despite the fact that recent advances by the Exomol project \citep{mckemmish_exomol_2019} have made the TiO line list sufficiently accurate to enable detections of TiO in ultrahot Jupiter atmospheres in some cases \citep{prinoth_titanium_2022}, predicted model-strengths tend to overestimate the true TiO line strength if it is present. In the case of the TiO detection in the transmission spectrum of WASP-189 b by \citet{prinoth_titanium_2022} and \citet{prinoth_time-resolved_2023}, the injected signal was approximately five or six times larger than the detected signal, which we attributed to the remaining inaccuracies in the TiO line list. \\
Figure\,\ref{fig:tio} shows that even though the noise level is down to the ppm-level, the injected signal suggests only a marginally significant detection if accounting for the five to six times over-predicted signal strength. Assuming that all Ti-bearing species are similarly affected by the depletion through the partial cold trap implies that even if TiO becomes successfully vertically remixed into the upper layers of the atmosphere, the abundance may be too low to allow a detection using current line lists. 
Sensitive observations with in particular JWST \citep{Mikal-Evans2021} will allow us to probe the band heads instead of individual lines, and could reveal the presence or absence of TiO in \target's atmosphere \citep{hoeijmakers_mantis_2024}. 

\section{Conclusions}
\label{sec:conclusion}
Thanks to the high S/N achieved, we successfully detect and confirm the presence of \alldetectionsccf in the atmosphere of \target using the cross-correlation technique. 
Additionally, we detect \alldetectionsnarrowband in resolved single lines, with a more detailed discussion of \ion{H}{$\alpha$} and \ion{Na}{I} presented in \citet{seidel_sub}.

Most notably, we detect the absorption of \Ti at high significance ($\sim19\sigma$ in the planetary rest frame). When combined with the detection of \V and compared to models, these findings confirm the previously observed depletion of \Ti-bearing species. Our results suggest that the cold-trapping mechanism responsible for this depletion exhibits a smooth transition over a range of equilibrium temperatures, with high S/N being crucial for detecting even the depleted species in the atmospheres of hot exoplanets.

Our observations of \target suggest limited mixing between the equatorial jet and mid-latitudes, where Ti material remains confined to equatorial regions, which is in contrast with models that predict tracer depletion at the equator when condensable molecules are included, highlighting the need for further investigation into atmospheric mixing.

We also discuss the non-detection of TiO in the context of line-list inaccuracies, depletion, and remixing processes. Our findings indicate that even if TiO is present, albeit depleted and remixed into the upper layers of the atmosphere, the current noise floor (at the ppm level) combined with inaccuracies in the line list may prevent its detection from the ground. Future studies using JWST to probe the band heads will be essential in determining the presence or absence of TiO with greater certainty. Our non-detection of TiO underscores the importance of synergies with space-based observatories, particularly JWST, as the only means to unravel some of the mysteries that remain elusive from ground-based observations.

The 4-UT mode of ESPRESSO, with its effective photon-collecting area equivalent to that of a 16 meter class telescope, serves as a valuable test-bed for pushing the limits of S/N on relatively faint targets. The present analysis also allows us to anticipate the observational capabilities of the soon-to-be-commissioned ELT, particularly with regard to time-resolved studies of exoplanet atmospheres. 

\begin{acknowledgements}
The authors acknowledge the ESPRESSO project team for its effort and dedication in building the ESPRESSO instrument. This work relied on observations collected at the European Organisation for Astronomical Research in the Southern Hemisphere. BP acknowledges financial support from the Walter Gyllenberg Foundation. JVS acknowledges support from the Munich Institute for Astro-, Particle and BioPhysics (MIAPbP) which is funded by the Deutsche Forschungsgemeinschaft (DFG, German Research Foundation) under Germany´s Excellence Strategy – EXC-2094 – 390783311. This work was partially funded by the French National Research Agency (ANR) project EXOWINDS (ANR-23-CE31-0001-01). RA acknowledges the Swiss National Science Foundation (SNSF) support under the Post-Doc Mobility grant P500PT\_222212 and the support of the Institut Trottier de Recherche sur les Exoplanètes (iREx). HMT acknowledges support from the “Tecnologías avanzadas para la exploración de universo y sus componentes" (PR47/21 TAU) project funded by Comunidad de Madrid, by the Recovery, Transformation and Resilience Plan from the Spanish State, and by NextGenerationEU from the European Union through the Recovery and Resilience Facility. ASM acknowledges financial support from the Spanish Ministry of 
Science and Innovation (MICINN) project PID2020-117493GB-I00 and from the Government of the Canary Islands project ProID2020010129. NMS acknowledges funding by the European Union (ERC, FIERCE, 101052347). Views and opinions expressed are however those of the author(s) only and do not necessarily reflect those of the European Union or the European Research Council. Neither the European Union nor the granting authority can be held responsible for them. This work was supported by FCT - Fundação para a Ciência e a Tecnologia through national funds and by FEDER through COMPETE2020 - Programa Operacional Competitividade e Internacionalização by these grants: UIDB/04434/2020; UIDP/04434/2020. MS acknowledges financial support from the Swiss National Science Foundation (SNSF) for project 200021\_200726. This project has also been carried out in the frame of the National Centre for Competence in Research PlanetS supported by the SNSF. FB acknowledges support from Bando Ricerca Fondamentale INAF 2023. This work makes use of \texttt{Astropy} \citep{astropy:2013,astropy:2018,astropy:2022}, \texttt{Matplotlib} \citep{matplotlib:2007}, \texttt{Numpy} \citep{harris2020array}, \texttt{lmfit} \citep{newville_2015_11813}, JAX \citep{jax2018github}, \texttt{NumPyro} \citep{phan_composable_2019}, \texttt{PyMultiNest} \cite{feroz_multinest_2011,buchner_x-ray_2014}, \texttt{SciPy} \citep{2020SciPy-NMeth}, and \texttt{seaborn} \citep{Waskom2021}.

\end{acknowledgements}

\bibliographystyle{aa} 
\bibliography{references}

{\clearpage
\begin{appendix}
\onecolumn
\section{Narrow-band resolved detections}
\begin{table*}[h!]
    \caption{Narrow-band resolved detections}\vspace{-1em}
    \begin{center}
            \begin{tabular}{cc|cccc}
                    \toprule
                    Element & $\lambda_\mathrm{air}$  & Contrast ($\sigma$) & FWHM & Shift & Data  \\
                      & $[\AA]$   & [$\%$]  & [$\AA$] & [$\si{\km\per\second}$] & \\
                    \midrule
                    \midrule
                non-metals &  & & & &  \\ 
                    \midrule
                    \midrule
                \ion{H}{$\alpha$} & $6562.81$ & $1.595\pm0.041~(39.1)$ & $1.13\pm0.03$ & $-0.25\pm1.07$ &  ingress\\
                                 &  & $1.70\pm0.048~(35.4)$ & $0.90 \pm 0.04$ & $-3.9 \pm 0.6$ & \citet{borsa_atmospheric_2021}\\
                \ion{H}{$\beta$}  & $4861.33$ & $0.543\pm0.058~(9.4)$& $0.67\pm0.05$ & $3.90\pm0.57$ & ingress \\
                 &  & $0.35 \pm 0.11~(3.2)$& $0.42 \pm 0.24$ & $-6.3 \pm 4.3$ & \citet{borsa_atmospheric_2021} \\
                \ion{H}{$\gamma$}  & $4340.47$ & -& -& -& combined \\
                    \midrule
                    \midrule
                Alkali metals &  & & & &  \\ 
                    \midrule
                    \midrule
                 \ion{Li}{} & $6707.76$ & $0.350\pm0.057~(6.1)$ & $0.12\pm0.07$ & $-2.23\pm0.44$ & ingress\\
                  & & $0.23 \pm 0.04~(5.8)$ & $0.43 \pm 0.07$ & $-2.7 \pm 1.6$ & \citet{borsa_atmospheric_2021}\\
                \midrule
                \ion{Na}{} & $5889.95$ & $0.583\pm0.043~(13.6)$ & $0.47\pm0.04$ & $-2.09\pm0.39$ &  ingress\\
                & & $0.447 \pm 0.045~(9.9)$ & $0.31 \pm 0.04$ & $-5.9 \pm 0.8$ &  \citet{borsa_atmospheric_2021}\\
                 & $5895.92$ & $0.551\pm0.040~(13.9)$ & $0.59\pm0.05$ & $-2.24\pm0.53$ & ingress \\
                 & & $0.51 \pm 0.048~(10.6)$ & $0.52 \pm 0.06$ & $-4.9 \pm 1.1$ & \citet{borsa_atmospheric_2021} \\
                \midrule
                 \ion{K}{} & $7698.96$ & $0.189\pm0.052~(3.63)$ & $0.23\pm 0.27$ & $-2.37\pm0.78$ & ingress \\
                 & & $0.23 \pm 0.05~(4.6)$ & $0.32 \pm 0.14$ & $-4.9 \pm 1.2$ & \citet{borsa_atmospheric_2021} \\
                    \midrule
                    \midrule
                Alkaline earth metals &  & & & &  \\ 
                    \midrule
                    \midrule
                 \ion{Mg}{I} & $4571.10$ & $0.381\pm0.057~(6.6)$& $0.25\pm0.06$&  $-5.78\pm3.86$ &  ingress \\
                 & & $0.40 \pm 0.08~(5.0)$& $0.17 \pm 0.03$&  $-7.9 \pm 1.0$ &  \citet{borsa_atmospheric_2021} \\
                \midrule
                  & $5167.32$ & $0.447\pm0.078~(5.7)$& $0.74\pm0.10$&  $-4.79\pm2.61$ & \\
                  \ion{Mg}{I}  & $5172.68$ & $0.313\pm0.055~(5.7)$& $0.80\pm0.11$&  $-1.88\pm0.49$ & ingress \\
                    & $5183.60$ & $0.472\pm0.045~(10.4)$& $0.79\pm0.06$&  $-1.47\pm0.78$ & \\
                    & $5167.32$ & & & & \\
                  \ion{Mg}{I}  & $5172.68$ & $-0.24 \pm 0.06~(4.0)$ & $13.4 \pm 4.4$ & $-6.3 \pm 1.8$ & \citet{borsa_atmospheric_2021} \\
                   (combined triplet) & $5183.60$ & & in \si{\km\per\second} & & \\
                 \midrule            
                  \ion{Ca}{II} K & $3933.66$ & $3.111\pm0.245~(12.7)$& $1.35\pm0.06$&  $-8.9\pm1.33$ & ingress \\
                  & & $5.22 \pm 0.17~(30.7)$& $1.12 \pm 0.04$&  $-17.6 \pm 1.4$ & \citet{borsa_atmospheric_2021} \\          
                  \ion{Ca}{II} H & $3968.47$ & $2.814\pm0.307~(9.2)$& $0.79\pm0.05$&  $-6.88\pm1.45$ & ingress \\
                   & & $4.18 \pm 0.19~(22.0)$& $0.95 \pm 0.05$&  $-19.0 \pm 1.6$ & \citet{borsa_atmospheric_2021} \\
                \midrule            
                  \ion{Sr}{I}  & $4607.33$ & $0.327\pm0.041~(8.1)$& $0.17\pm0.03$ & $-2.16\pm0.57$ & ingress \\
                  \ion{Sr}{I}  & $4607.33$ & $0.120\pm0.040~(3.0)$& $0.24\pm0.12$ & $-4.17\pm1.00$ & egress \\
                      \midrule
                \ion{Sr}{II}  & $4077.71$ & $0.687\pm0.067~(10.2) $& $0.28\pm0.08$ & $-1.42\pm0.24$ & combined \\
                  & $4215.52$ & $0.413\pm0.044~(9.4) $& $0.30\pm0.04$ & $-1.82\pm1.06$ & combined \\
                     \midrule
                \ion{Ba}{II}  & $4554.03$ & $0.322\pm0.073~(4.4)$ & $0.13\pm0.05$& $-2.05\pm0.33$ & ingress \\
                    \midrule
                \ion{Ra}{II}  & $3814.40$ & - & -& - & ingress \\
                & $4682.24$ & - & -& - & ingress \\
                    \midrule
                    \midrule
                Transition metals &  & & & &  \\ 
                    \midrule
                    \midrule
                  & $4030.76$ & & &  \\
                  \ion{Mn}{}  & $4033.07$ &$0.290\pm0.039~(7.4) $& $17.09\pm3.37$ & $-6.18\pm01.01$ & ingress \\
                  (combined triplet)  & $4034.49$ & & in $ \si{\km\per\second} $ & & \\
                  & $4030.76$ & & &  \\
                  \ion{Mn}{}  & $4033.07$ &$0.29 \pm 0.089~(3.6) $& $15.3 \pm 3.9$ & $-5.2 \pm 2.1$ & \citet{borsa_atmospheric_2021} \\
                  (combined triplet)  & $4034.49$ & & in $ \si{\km\per\second} $ & & \\
                  \bottomrule
            \end{tabular}
    \end{center} 
    \textit{Note:} The detections from \citet{borsa_atmospheric_2021} are given as a reference. If missing, the element was not previously detected. The FWHM in \citet{borsa_atmospheric_2021} were provided in $\si{\km\per\second}$ and were converted here to \AA\, for comparability. 
    \label{tab:narrowdetections}
\end{table*}
\newpage

\twocolumn
\section{Narrowband detection of \ion{Ba}{II}}
Fig.\,\ref{fig:BaII_narrowband} shows the tentative \ion{Ba}{II} detection for the ingress dataset at $\lambda =$ \SI{4554.03}{\AA}.
\begin{figure}[h!]
    \centering
    \includegraphics[width=0.7\linewidth]{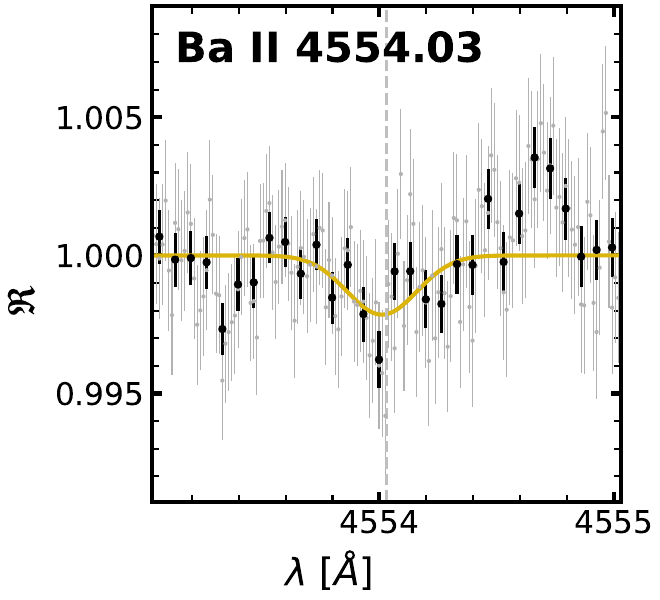}
    \caption{Tentative \ion{Ba}{II} detection for the ingress dataset at $\lambda =$ \SI{4554.03}{\AA}. The original data are shown in grey, binned by a factor of 5 for better visibility in black. The Gaussian fit is shown in a golden, solid line.}
    \label{fig:BaII_narrowband}
\end{figure}

\section{True ingress and egress corner plots}
\label{app:true_limbs}
We performed a Gaussian fit using JAX \citep{jax2018github,bingham2018,phan_composable_2019} on the true ingress and egress, which correspond to the three exposures at the beginning and end of the transit, where only the leading and trailing limb is in front of the star, respectively. In principle, absorption occurring only at one limb could mimic a depletion when averaged over the entire transit. However, our fit shows that absorption originates from both limbs (see Fig.\,\ref{fig:true_limbs}).

\begin{figure}[h!]
    \centering
    \includegraphics[width=\linewidth]{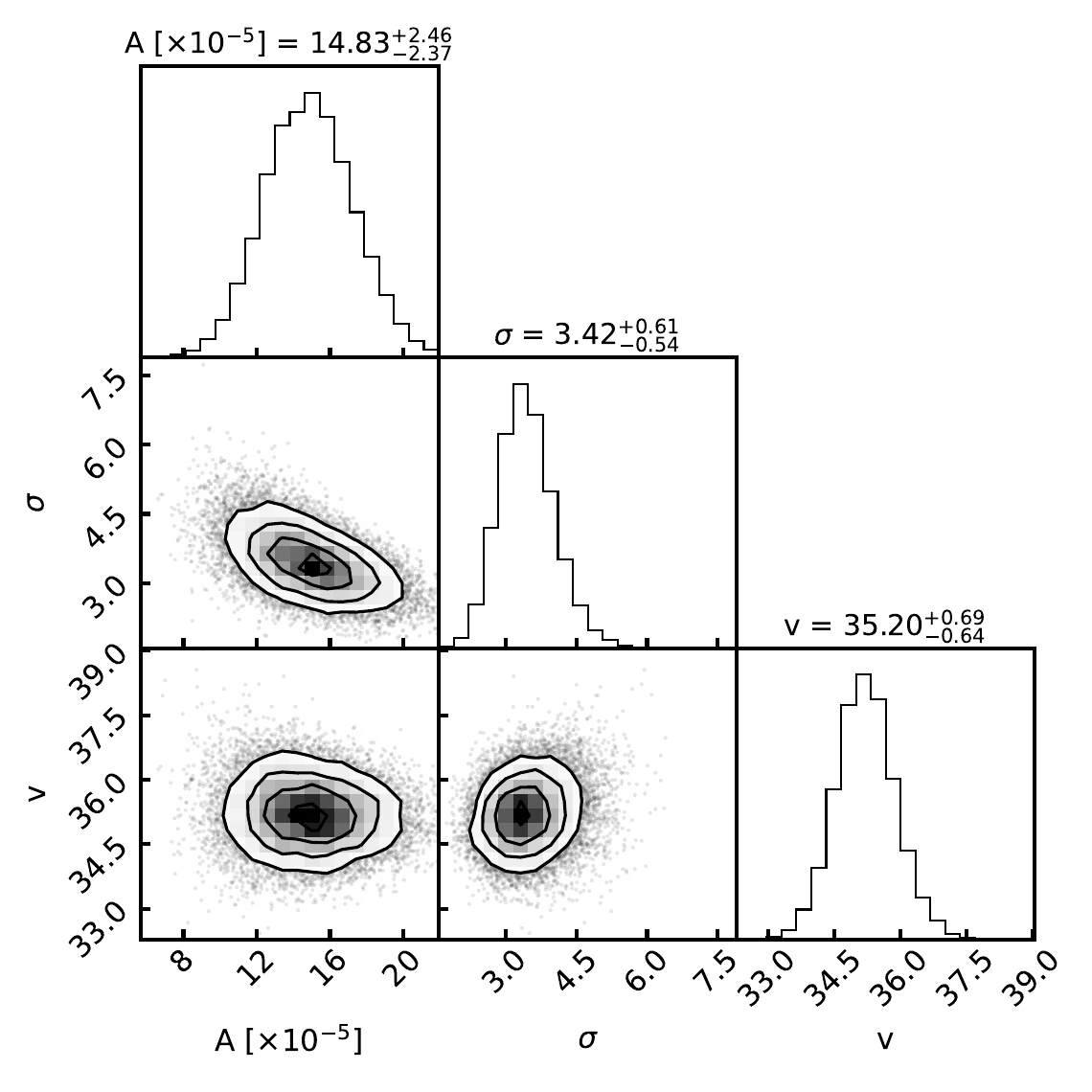}
    \includegraphics[width=\linewidth]{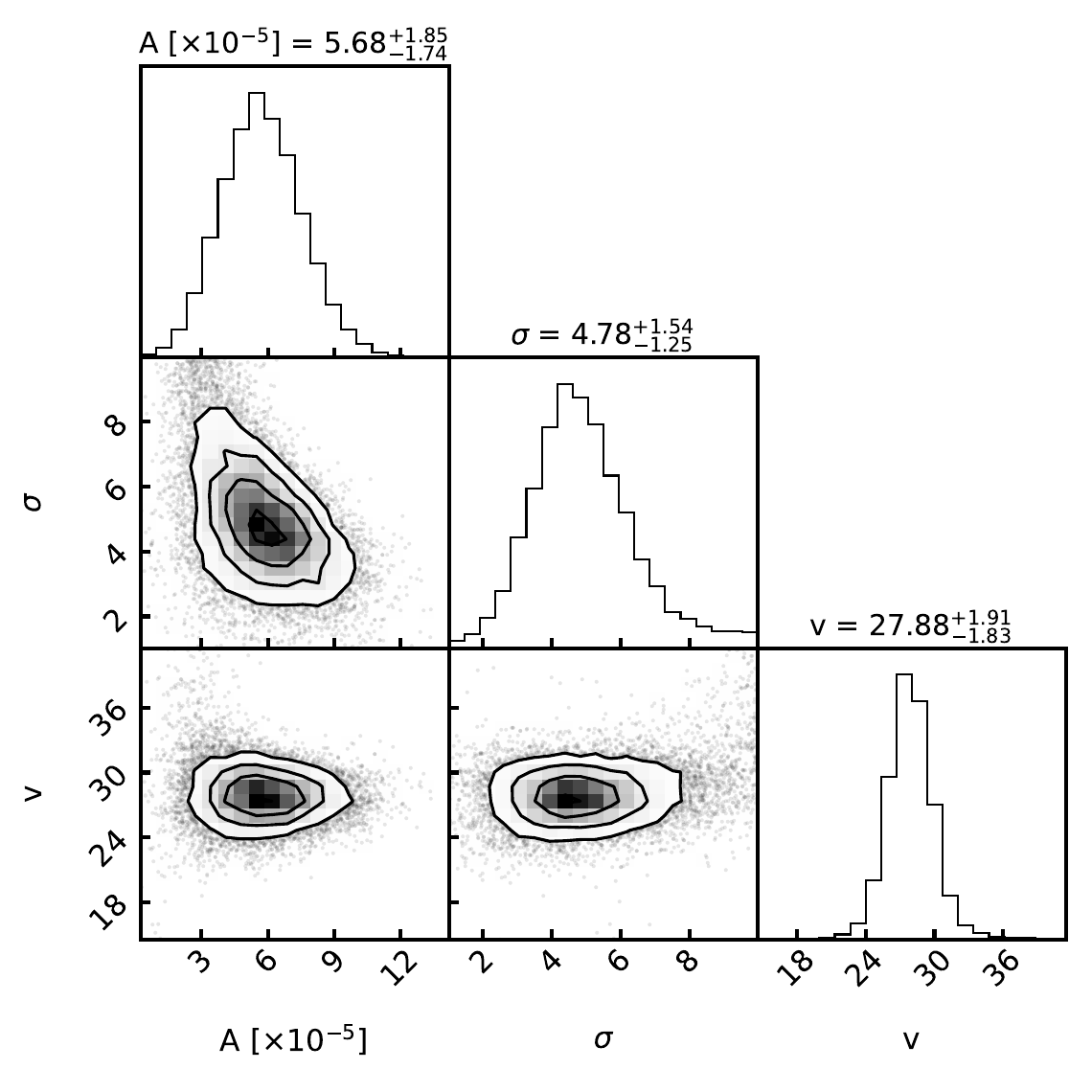}
    \caption{True limb absorption (three exposures each) of \Ti for ingress (top) and egress (bottom). $A$ is the absorption depth (Gaussian) in parts per ten-thousand ($\times 10^{-5}$), $s$ is the Gaussian width (not FWHM) in \si{\km\per\second} and $v$ is the centre velocity of the Gaussian in \si{\km\per\second}. The rest frame velocity is \SI{38.36 \pm 0.06}{\km\per\second} for reference.}
    \label{fig:true_limbs}
\end{figure}
\end{appendix}}

\label{LastPage}

\end{document}